\documentclass[usenatbib]{mn2e}

\usepackage{apjfonts,amsfonts,verbatim}
\usepackage{bm}
\usepackage{mathtools}
\usepackage{xcolor}
\usepackage{natbib}
\usepackage{lscape}
\usepackage{afterpage}  
\usepackage{booktabs}
\usepackage{tabularx}

\usepackage{threeparttable}

\usepackage{graphicx}	
\usepackage{amsmath}	
\usepackage{amssymb}	

\newcommand{\lya}{Ly$\alpha$}
\newcommand{\ha}{H$\alpha$}
\newcommand{\hb}{H$\beta$}
\newcommand{\oiii}{[\rm O\,{\textsc {iii}}]}

\newcommand{\HI}{\rm H\,{\textsc {i}}}
\newcommand{\HII}{\rm H\,{\textsc {ii}}}
\newcommand{\angstrom}{\textup{\AA}}

\newcommand{\kms}{\,\ifmmode{\mathrm{km}\,\mathrm{s}^{-1}}\else km\,s${}^{-1}$\fi}

\title[Gas Kinematics in LAB1]{Revisiting the Gas Kinematics in SSA22 Lyman-$\alpha$ Blob 1 with Radiative Transfer Modeling in a Multiphase, Clumpy Medium}

\author[Li et al.]{
Zhihui Li,$^{1}$\thanks{E-mail: zhihui@caltech.edu}
Charles C. Steidel,$^{1}$
Max Gronke$^{2,3}$\thanks{Hubble fellow}
and Yuguang Chen$^{1}$
\\
$^{1}$Cahill Center for Astrophysics, California Institute of Technology, MC 249-17, 1200 East California Boulevard, Pasadena, CA 91125, USA\\
$^{2}$Department of Physics, University of California, Santa Barbara, CA 93106, USA\\
$^{3}$Department of Physics \& Astronomy, Johns Hopkins University, Baltimore, MD 21218, USA
}

\date{}

\begin{document}
\label{firstpage}
\pagerange{\pageref{firstpage}--\pageref{lastpage}}
\maketitle

\begin{abstract}
We present new observations of Lyman-$\alpha$ (\lya) Blob 1 (LAB1) in the SSA22 protocluster region ($z$ = 3.09) using the Keck Cosmic Web Imager (KCWI) and Keck Multi-object Spectrometer for Infrared Exploration (MOSFIRE). We have created a narrow-band \lya\ image and identified several prominent features. By comparing the spatial distributions and intensities of \lya\ and \hb, we find that recombination of photo-ionized \HI\ gas followed by resonant scattering is sufficient to explain all the observed \lya/\hb\ ratios. We further decode the spatially-resolved \lya\ profiles using both moment maps and radiative transfer modeling. By fitting a set of multiphase, `clumpy' models to the observed \lya\ profiles, we manage to reasonably constrain many parameters, namely the \HI\ number density in the inter-clump medium (ICM), the cloud volume filling factor, the random velocity and outflow velocity of the clumps, the \HI\ outflow velocity of the ICM and the local systemic redshift. Our model has successfully reproduced the diverse \lya\ morphologies, and the main results are: (1) The observed \lya\ spectra require relatively few clumps per line-of-sight as they have significant fluxes at the line center; (2) The velocity dispersion of the clumps yields a significant broadening of the spectra as observed; (3) The clump bulk outflow can also cause additional broadening if the \HI\ in the ICM is optically thick; (4) The \HI\ in the ICM is responsible for the absorption feature close to the \lya\ line center.
\end{abstract}


\section{Introduction}

\lya\ Blobs (LABs) -- spatially extended (projected sizes $\gtrsim$ 100 kpc) gaseous nebulae at high redshift ($z \gtrsim$ 2) with immense \lya\ luminosities ($L_{\rm Ly\alpha} \sim 10^{43-44}$\,erg\,$\rm s^{-1}$) -- are among the most enigmatic and intriguing objects in the universe. To date, hundreds of LABs have been discovered (e.g., \citealt{Francis96, Fynbo99, Keel99, Steidel2000, Matsuda04, Matsuda11, Dey05, Saito06, SmithJarvis07, Hennawi09, Ouchi09, Prescott09, Prescott12, Erb11, Cai17}), yet their physical origin remains murky. Many of the LABs have been found in overdense regions associated with massive proto-clusters, which will presumably evolve into rich galaxy clusters observed today (e.g., \citealt{Steidel98, Prescott08, Yang09, Yang10, Hine16}). Hence, the study of LABs may elucidate the formation process of massive galaxies and the mechanisms of concurrent feedback events.

What are the possible energy sources that power the observed \lya\ emission of LABs? Thus far, numerous attempts have been made to answer this fundamental question, but a consensus is yet to be reached. Among many proposed scenarios, one of the most plausible \lya\ production mechanisms is photo-ionization via embedded energetic sources (e.g., starburst galaxies or AGNs) followed by subsequent recombination \citep{Haiman01, Cantalupo05, Cantalupo14}. This scenario has been corroborated by the discovery of luminous galaxies and AGNs \citep{Chapman01, Dey05, Geach05, Geach07, Geach09, Colbert06, Webb09} inside some LABs via infrared and submillimeter observations. If the ionizing sources are starbursts, supernova-induced energetic winds may be triggered \citep{Heckman90, Taniguchi00, Taniguchi01, Mori04}, producing outflowing super-bubbles and additional \lya\ emission via shock heating. Evidence for the existence of such `superwinds' includes the observed double-peaked \lya\ profiles \citep{Ohyama03} and bubble-like structures \citep{Matsuda04}. Alternatively, \lya\ emission can originate from cooling radiation via accretion of cold gas streams in dark matter halos onto protogalaxies \citep{Haiman00, Fardal01, Furlanetto05, Dijkstra06a, Dijkstra06b, Scarlata09, Goerdt10, Faucher10, Rosdahl12}. This explanation is especially favored for LABs with no or only weak associated energy sources identified even with deep multi-wavelength observations \citep{Nilsson06, SmithJarvis07, Saito08, Smith08}. In either case, a substantial fraction of the \lya\ photons will be resonantly scattered multiple times before escape \citep{Steidel10, Steidel11}, although the `cold accretion' scenario is supposed to induce a lower degree of polarization due to a lower chance of scattering from the inside out \citep{Dijkstra09, Hayes11, Trebitsch16, Eide2018}.

In this paper, we present new observations and analyses of one of the first LABs ever discovered, SSA22-Blob1 (LAB1, \citealt{Steidel2000}). LAB1 is one of the brightest and largest LABs discovered to date, with a \lya\ luminosity of $\sim$\,1.1\,$\times$\,10$^{44}$\,erg\,s$^{-1}$ \citep{Weijmans10} and a spatial extent of $\sim$\,100\,kpc \citep{Matsuda04}. Since its discovery, LAB1 has been studied extensively, at wavelengths including X-ray \citep{Geach09}, optical \citep{Ohyama03, Bower04, Weijmans10}, infrared (IR, \citealt{Uchimoto08, Uchimoto12, Webb09}) and submillimeter (submm, \citealt{Geach05, Matsuda07, Geach14, Hine16}). Two Lyman-break galaxies (LBG), C11 and C15 \citep{Steidel2000, Steidel03, Matsuda04}, and multiple dust-obscured star-forming galaxies \citep{Geach07, Geach14, Geach16} have been identified within LAB1. However, X-ray observations yield non-detections, indicating the absence of (Compton-thin) AGNs \citep{Geach09}.

To determine the principle energy source(s) powering LAB1, three main approaches have been adopted: the first is to infer the gas kinematics (e.g., inflows v.s. outflows) from the observed properties of \lya\ as well as other non-resonant emission lines (e.g., \oiii, \ha, \hb). For example, \citet{Bower04} and \citet{Weijmans10} measured a velocity shear of the \lya\ emission from C11 and C15 using integral-field spectroscopy, which suggests the presence of outflows. On the other hand, \citet{McLinden13} reported a nearly zero velocity offset between \lya\ and \oiii\ in C11 and C15, which they interpreted as an absence of strong outflows. An alternative approach is to compare the available energy budget of possible energy sources with the observed \lya\ emission. For example, \citet{Geach16} deduced the IR luminosities and corresponding star formation rate (SFR, $\sim$\,150\,$M_{\odot}$\,yr$^{-1}$) of the embedded sources from their 850\,$\mu$m flux density measured with ALMA, and found that this energy budget is sufficient to power the observed \lya\ luminosity. However, as it is difficult to independently constrain the fraction of \lya\ photons that escape from the galaxy and scatter into our line of sight, additional 
energy sources (e.g., cold accretion) cannot be ruled out entirely \citep{Geach14, Hine16}. Thirdly, \citet{Hayes11} and \citet{Beck16} have measured polarized \lya\ emission using polarimetric imaging. Although they claimed that this result should be strongly supportive of a `central powering + scattering' model, \citet{Trebitsch16} pointed out that the scattering inside the cold filaments in the `cold accretion' scenario could still account for the degree of polarization observed.

In this paper, we use an advanced kinematic approach to further test the feasibility of the `central powering + scattering' scenario. The traditional kinematic approach -- inferring the underlying gas velocity field from the observed peak shifts and line widths (e.g., \citealt{McLinden13}) is worth scrutinizing, as resonant scattering may modify the line profiles in a very complex way. Instead, we model the \lya\ profiles using Monte-Carlo radiative transfer (MCRT). Due to its computationally expensive nature, \lya\ MCRT modeling normally assumes a simple, idealized geometry, e.g., a spherically symmetric expanding shell of \HI\ gas surrounding a central \lya\ emitting source (the `shell model', \citealt{Verhamme06, Dijkstra06b}). This simple model has successfully reproduced many observed \lya\ spectra (e.g., \citealt{Schaerer08, Verhamme08, Dessauges10, Vanzella10,Gronke17}), although it has also encountered some challenges for those with multiple peaks (\citealt{Verhamme08, Kulas12, Thorsen2017}) or very large line widths (\citealt{Hashimoto15, Yang16, Yang17a, Orlitova18}). Moreover, recent observations have shown increasing evidence that the circumgalactic medium (CGM), just like the interstellar medium (ISM), is multiphase and clumpy (e.g., the \lya\ emission and metal absorption line observations of high-redshift quasars (\citealt{Cantalupo14, Hennawi15})), which is further corroborated by simulations with increased spatial resolution (e.g., \citealt{Hummels19}). Therefore, a more realistic model that accounts for the multiphase nature and clumpy geometry of \HI\ gas is needed to properly characterize the radiative transfer processes of \lya\ photons.

Up to now, this multiphase `clumpy' model has been explored theoretically via both semi-analytical calculations \citep{Neufeld91} and Monte-Carlo simulations \citep{Hansen06, Dijkstra12, Laursen13, Gronke16_model}. However, due to its complex and multivariate nature, the multiphase `clumpy' model has not been widely used in fitting real \lya\ spectra (albeit the first attempt made in \citealt{Forero18}). In this work, we use the framework proposed by \citet{Gronke16_model} to model the spatially-resolved \lya\ spectra in LAB1.

In addition to the \lya\ observations in the optical (rest-frame UV) using the Keck Cosmic Web Imager (KCWI; \citealt{2010SPIE.7735E..0MM, 2012SPIE.8446E..13M}), we have carried out near-infrared (NIR, rest-frame optical) spectroscopic observations using the Keck Multi-object Spectrometer for Infrared Exploration (MOSFIRE; \citealt{McLean10, McLean12}). By comparing the spatial distribution of \lya, \oiii\ and \hb\ emission and fitting \lya\ line profiles, we map the kinematic structure of \HI\ in LAB1 and constrain its possible powering mechanism(s). 

The structure of this paper is as follows. In \S\ref{sec:data}, we describe our KCWI and MOSFIRE observations and data reduction procedures. In \S\ref{sec:kinematic}, we present our new observational results and analyses. In \S\ref{sec:RT}, we detail the methodology and present our results of radiative transfer modeling using the multiphase, clumpy model. In \S\ref{sec:conclusion}, we summarize and conclude. Throughout this paper we adopt a flat $\Lambda$CDM cosmology with $\Omega_{\rm m}$ = 0.315, $\Omega_{\Lambda}$ = 0.685, and $H_{0}$ = 67.4 km s$^{-1}$ Mpc$^{-1}$ \citep{Planck18}. We use the following vacuum wavelengths: 1215.67\,\AA\ for \lya, 4862.683\,\AA\ for H$\beta$, and 4960.295/5008.240\,\AA\ for \oiii\ from the Atomic Line List v2.04\footnote{http://www.pa.uky.edu/$\sim$peter/atomic/index.html}.

\section{Observations and Data Reduction}\label{sec:data}

\subsection{KCWI Observations}\label{sec:KCWIdata}
The KCWI observations of LAB1 were carried out on the night of 2018 June 16, with a seeing of $\sim$ 1.0$\arcsec$ full width at half maximum (FWHM). We used the KCWI large slicer, which provides a contiguous field-of-view (FOV) of 20.4$\arcsec$ (slice length) $\times$ 33$\arcsec$ (24 $\times$ 1.35\arcsec\ slice width). With the BM VPH grating set up for $\lambda_{\rm c} = 4800\,$\AA, the wavelength coverage is $\sim$\,4260 - 5330\,$\rm \angstrom$, with spectral resolution $R\simeq 1800-2200$. The data were obtained as 9 individual 1200\,s exposures, with small telescope offsets in the direction perpendicular to slices applied between each, in an effort to recover some spatial resolution given the relatively large slice width. The total on-source exposure time was 3 hours. 

Individual exposures were reduced using the KCWI Data Reduction Pipeline\footnote{https://github.com/Keck-DataReductionPipelines/KcwiDRP}, which includes wavelength calibration, atmospheric refraction correction, background subtraction, and flux calibration. The individual datacubes were then spatially re-sampled onto a uniform astrometric grid with 0.3\arcsec\ by 0.3\arcsec\ spaxels, with a sampling of 0.5\,\AA\,pix$^{-1}$ (4.75 pixels per spectral resolution element) along the wavelength axis, using a variant of the `drizzle' algorithm (with a drizzle factor of 0.9) in the \texttt{MONTAGE}\footnote{http://montage.ipac.caltech.edu} package. The re-sampled cubes were then combined into a final stacked cube by averaging with exposure time weighting. Owing to the coarser spatial sampling in the long dimension of the spatial cube, the PSF in the final datacube is elongated along the N-S direction, with FWHM $\simeq 0.96\arcsec \times 1.44\arcsec$ (X-direction and Y-direction, respectively). 
 
The resampled final datacube covers a scientifically useful solid angle of 18.9$\arcsec \times 32.7\arcsec$ on the sky, and a wavelength range (vacuum, heliocentric) of 4214 - 5243\,\AA. A variance image with the same dimensions was created by propagating errors based on a noise model throughout the data reduction. 
 
\subsection{MOSFIRE Observations}\label{sec:MOSFIREdata}

We observed selected regions of LAB1, chosen to include the highest \lya\ surface brightness areas as determined from a very deep narrow-band \lya\ image (see \citealt{Steidel11,Nestor11}) using MOSFIRE \citep{McLean10, McLean12,Steidel14} on the Keck I telescope. Spectra in the near-IR $K$ band (1.95 - 2.40\,${\mu}$m) were obtained using four different slitmasks, each of which included a slit passing through part of LAB1 with a different RA, Dec, and position angle (PA). The four slits are labeled as `slit 1' through `slit 4' in Figure \ref{fig:Lya_images}, and the observations are summarised in Table \ref{tab:mosfire_obs}. Slits 1 - 3 were obtained using slits of width 0.7$\arcsec$, providing spectral resolving power of $R \simeq 3700$; slit 4 observations used a 1.0$\arcsec$ wide slit, yielding $R \simeq 2600$. The observations were obtained during four different observing runs between 2012 June and 2019 November, under clear skies with seeing in the range 0.43$\arcsec$ - 0.53$\arcsec$, as summarised in Table \ref{tab:mosfire_obs}. 

The MOSFIRE $K$ band observations (slit 1 was also observed in $H$ band) were all obtained using an ABAB nod pattern along the slit direction with nod amplitude of 3$\arcsec$ between position A and position B for slits 1, 2, and 3, and 15$\arcsec$ for slit 4. Total integration times were 1.5 - 4.0 hours, as listed in Table \ref{tab:mosfire_obs}, composed of 30 - 80 individual 180\,s exposures. The data for each observation sequence were reduced using the MOSFIRE data reduction pipeline\footnote{https://github.com/Keck-DataReductionPipelines/MosfireDRP} to produce two-dimensional, rectified, background-subtracted vacuum wavelength calibrated spectrograms (see \citealt{Steidel14} for details). Observations obtained on different observing nights using the same slitmask were reduced independently; the 2-D spectrograms were shifted into the heliocentric rest frame and combined with inverse variance weighting using tasks in the \texttt{MOSPEC} \citep{Strom17} analysis package.

\begin{table}
    \centering
    \scriptsize \caption{MOSFIRE $K$-band observations of LAB1.}
    \label{tab:mosfire_obs}
    \begin{tabular}{cccccccc}
    \hline\hline
    Name & Width & $R$ & PA & Exp & Seeing & Date of Obs & Nod  \\ 
     (1)  & (2) & (3) & (4) & (5) & (6) & (7) & (8) \\
    \hline
    Slit 1 & 0.7 & 3660 & $-68.0$ & 4.0 & 0.50 & 2012 Sep15 & $~3.0$\\
    Slit 2 &  0.7 & 3660 & $-3.5$ & 1.5 & 0.43 & 2012 Jun30 & $~3.0$ \\
    Slit 3 &  0.7 & 3660 & $~27.0$ & 1.5 & 0.52 & 2012 Sep13 & $~3.0$ \\
    Slit 4 &  1.0 & 2560 & $-54.0$ & 2.5 & 0.53 & 2019 Jun15 & $15.0$ \\
    \hline\hline
    \end{tabular}
    \begin{tablenotes}
    \item \textbf{Notes.} The details of the MOSFIRE $K$-band observations of LAB1. The columns are: (1) slit name; (2) slit width ($\arcsec$); (3) resolving power ($\lambda/\Delta\lambda$); (4) slit PA (degrees E of N); (5) exposure time in hours; (6) seeing FWHM ($\arcsec$); (7) UT date of observation; (8) nod amplitude between A and B positions ($\arcsec$).
    \end{tablenotes}
\end{table}

\section{The gas kinematic structure of LAB1}\label{sec:kinematic}

\subsection{Spatial Distribution of \lya\ Emission}\label{sec:spatial}

To get an overview of the \lya\ surface brightness (SB) distribution in LAB1, we first generate a \lya\ narrow-band image by optimally summing all the \lya\ fluxes over the relevant wavelength range. Here we follow the `matched filtering' procedures for creating a narrow-band image \citep{Herenz20} using \texttt{LSDCat} \citep{Herenz17}. Firstly, we apply spatial filtering to the continuum-subtracted KCWI datacube using a 2D Gaussian filter with a constant 1.2$\arcsec$ FWHM, which equals the seeing point spread function (PSF) measured from a bright star in the SSA22 field. 
Secondly, we apply a 1D Gaussian spectral filter with FWHM = 1000\,km\,s$^{-1}$, which is the typical observed \lya\ line width estimated via visual inspection. Thirdly, we use this filtered datacube to generate a S/N cube. We can then choose appropriate S/N thresholds for the filtered datacube to produce SB and moment maps (see Section \ref{sec:profiles}).

\begin{table}
    \centering
    \scriptsize \caption{Continuum sources identified in LAB1.}
	\label{tab:sources}
	\renewcommand{\arraystretch}{1.2}
	\begin{tabular}{cccccc} 
		\hline \hline
		Name & RA (J2000) & Dec (J2000) & $z_{\rm sys}$ & Type & Refs.\\
		\hline
		C11$^{\rm a}$ & 22:17:25.70 & +00:12:34.7 & 3.0980$\pm$0.0001 &\oiii\ & (1)(2)\\
		C15$^{\rm a}$ & 22:17:26.15 & +00:12:54.7 & 3.0975$\pm$0.0001 &\oiii\ & (1)(2)\\
		ALMA-a & 22:17:25.94 & +00:12:36.6 & ...& ...&(3)\\
		ALMA-b & 22:17:26.01 & +00:12:36.4 & ...& ...&(3)\\
		ALMA-c & 22:17:26.11 & +00:12:32.4 & 3.1000$\pm$0.0003 &{[\rm C\,{\textsc {ii}}]} &(5)\\
		       & 22:17:26.10 & +00:12:32.3 & 3.0993$\pm$0.0004 &\oiii\ &(3)\\
		K1 & 22:17:25.70 & +00:12:38.7 & 3.1007$\pm$0.0002& \oiii\ &(4)\\
		c$_1$ & 22:17:25.94 & +00:12:36.0 & 3.0988&\oiii\ &(1)\\
		c$_2$/S1 & 22:17:26.08 & +00:12:34.2 & 3.0968&\oiii\ &(1)(3)\\
		c$_3$ & 22:17:26.05 & +00:12:38.7 & 2.7542&\lya\ &(1)\\
		\hline \hline
	\end{tabular}
	\begin{tablenotes}
    \item $^{\rm a}$Originally defined in \citet{Steidel2000}. \\
    \textbf{References.} (1) This work; (2) \citet{McLinden13}; (3) \citet{Geach16}; (4) \citet{Kubo15}; (5) \citet{Umehata17}.
    \end{tablenotes}
\end{table}

\begin{figure*}
\centering
\includegraphics[width=0.547\textwidth]{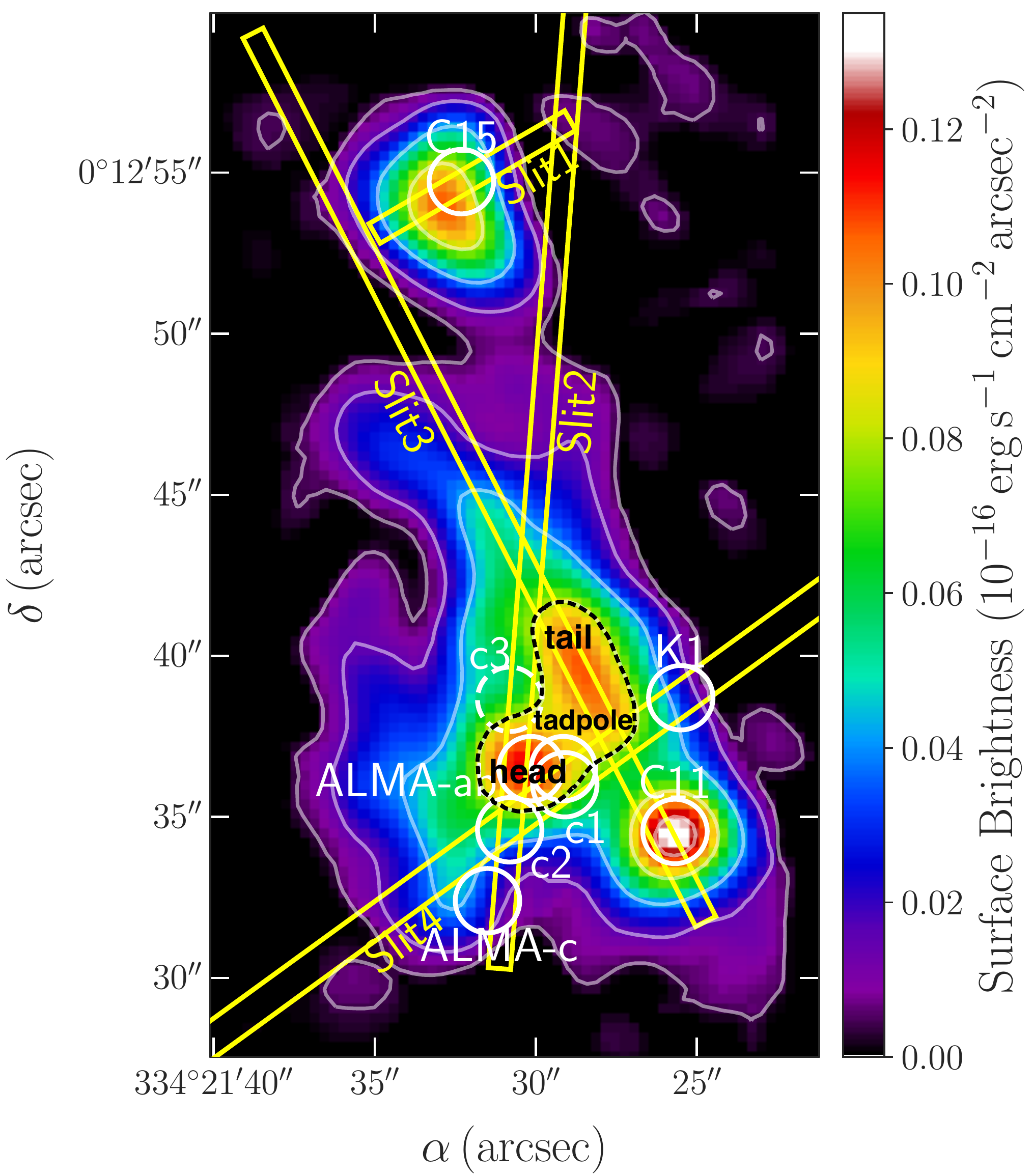}
\includegraphics[width=0.444\textwidth]{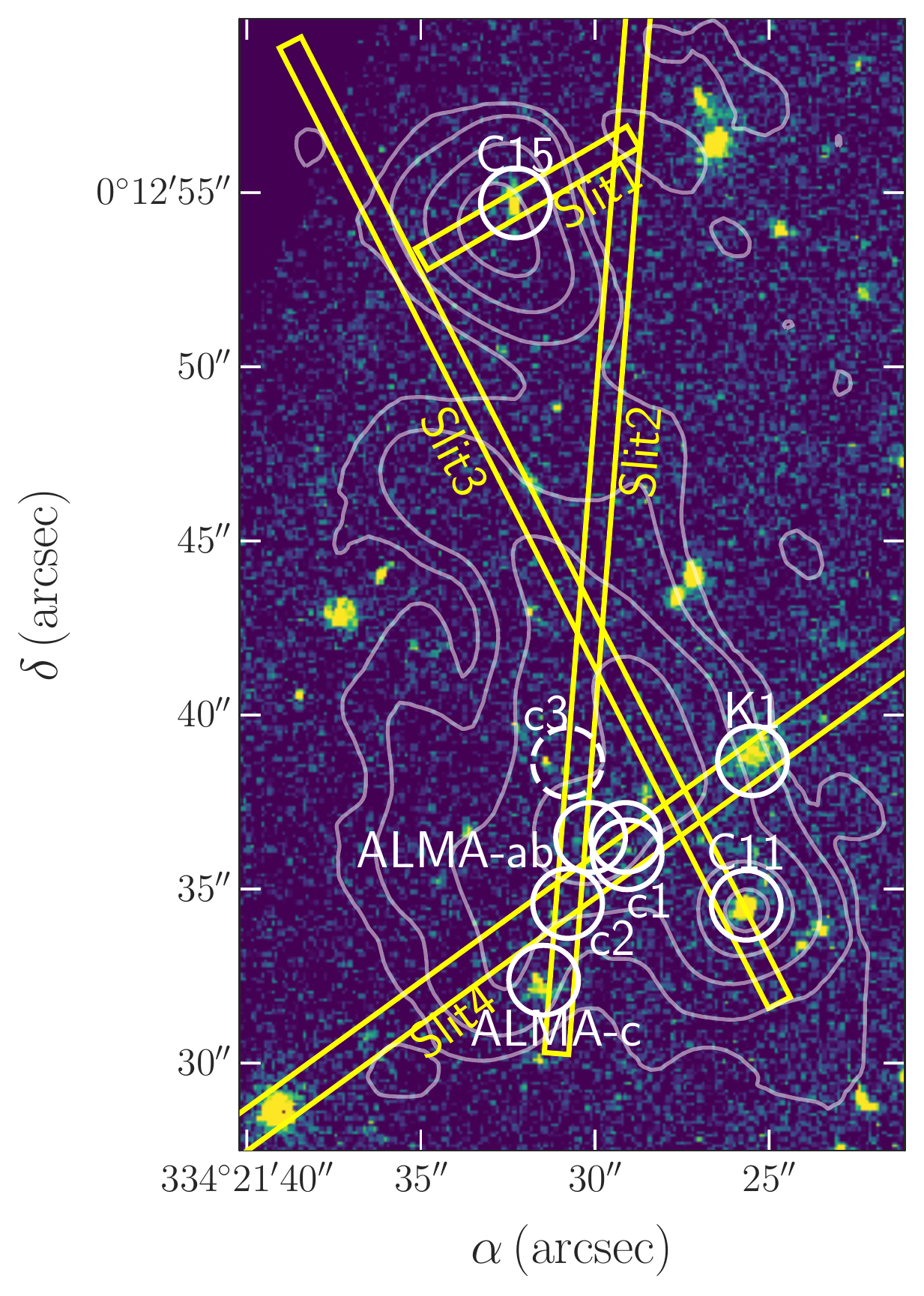}\\
    \caption{\lya\ and continuum images of LAB1. \emph{Left}: The narrow band \lya\ image, obtained by collapsing the original KCWI datacube over 4959 -- 5009\,\AA, which contains the \lya\ line (see Section 2). The UV continuum near the wavelength of \lya\ has been subtracted. \emph{Right}: The \emph{HST}/STIS optical continuum image. The positions of four MOSFIRE slits (Slit 1-4), two Lyman-break galaxies (C11 and C15), three submillimeter sources (ALMA-a, b, and c), a $K$-band selected galaxy (K1) and three \oiii/\lya\ serendipitous sources (c$_1$ to c$_3$) have been marked on each image (see Table \ref{tab:sources}). The \lya\ isophotes with levels of SB$_{\rm Ly\alpha}$ = [120,\,80,\,40,\,15,\,4]$\,\times\,10^{-19}$\,erg\,s$^{-1}$\,cm$^{-2}$\,arcsec$^{-2}$ have also been overlaid. All images have been registered to the same world-coordinate system.
    \label{fig:Lya_images}}
\end{figure*}

In the left panel of Figure \ref{fig:Lya_images}, we present our narrow-band \lya\ image. It is constructed by summing over all the voxels of the filtered datacube with S/N $\geq$ 4 over 4959 -- 5009\,\AA, which should enclose all possible \lya\ emission. To further examine whether the \lya\ emission coincides with the identified sources, we also present the \emph{HST}/STIS optical continuum image of LAB1\footnote{The KCWI and \emph{HST}/STIS images have been registered to the same world-coordinate system using cross-correlation.}.

We have also marked the positions of previously identified sources on each image as references. Among these sources, C11 and C15 are both LBGs \citep{Steidel2000}; ALMA-a, b and c are three submillimeter galaxies \citep{Geach16}; K1 is a $K$-band selected galaxy \citep{Kubo15}; c$_1$ and c$_2$ (the same as S1 in \citealt{Geach16}) are two \oiii\ serendipitous sources; c$_3$ is a \lya\ serendipitous source at a lower redshift ($z$ = 2.7542). The detailed information (especially spectroscopic redshifts, if available) of all the identified sources are presented in Table \ref{tab:sources}. The \lya\ isophotes (contours with the same SB) with levels of SB$_{\rm Ly\alpha}$ = [120,\,80,\,40,\,15,\,4]$\,\times\,10^{-19}$\,erg\,s$^{-1}$\,cm$^{-2}$\,arcsec$^{-2}$ have also been overlaid onto each image.

Several prominent features are evident in Figure \ref{fig:Lya_images}: (1) In general, the regions with the highest SB are associated with identified sources (e.g., C11, C15 and ALMA-a), although the position of the maximum \lya\ SB may be offset from the continuum source (e.g., C15); (2) An exception worth noting is a tadpole-shaped structure (marked in Figure \ref{fig:Lya_images}), which starts from the ALMA-ab sources, wriggles towards the north-west first and then north-east. Interestingly, although the `head' of the tadpole overlaps with ALMA-a, its `tail' does not overlap with any source; (3) The regions with identified continuum sources do not necessarily have significant \lya\ emission (e.g., ALMA-c, c$_2$, K1).

\subsection{Spatial Distribution of [O\,{\textsc {III}}] and \hb\ Emission}\label{sec:spatial_oiii}

To test whether the extended \lya\ emission is produced `in situ' or `ex situ' (the latter requires scattering), we further use MOSFIRE to map the spatial distribution of two other non-resonant lines, \oiii\ and \hb, and quantitatively compare them with \lya\ emission at the same spatial position. The positions of the four MOSFIRE slits are also shown in Figure \ref{fig:Lya_images}. 

\begin{figure}
\includegraphics[width=0.48\textwidth]{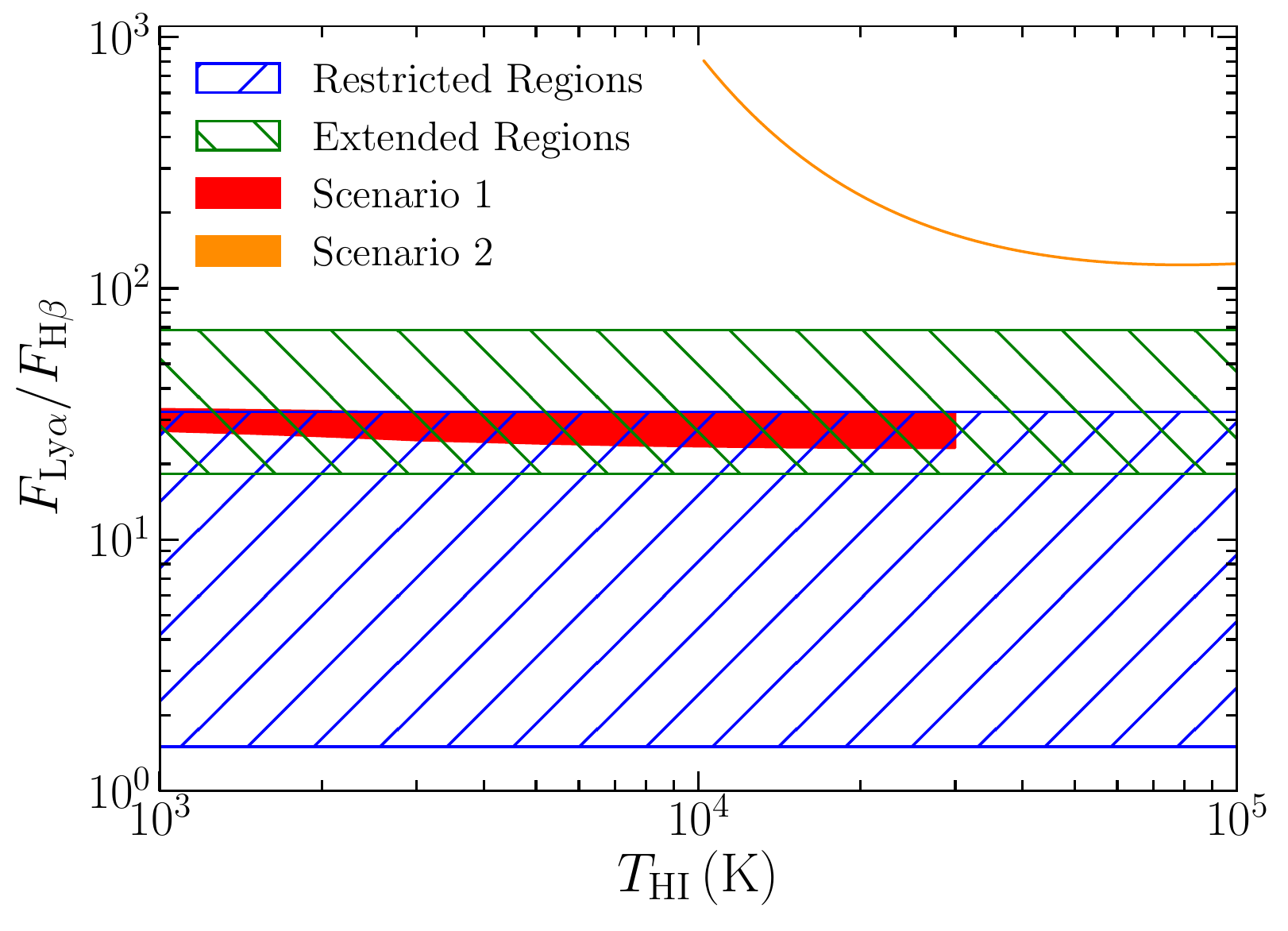}
    \caption{$F_{\rm Ly\alpha}$/$F_{\rm H\beta}$ as a function of the \HI\ gas temperature $T_{\rm \HI}$ in the photo-ionization (red patch) and collisional excitation (orange curve) scenarios. Also shown are the ranges of $F_{\rm Ly\alpha}$/$F_{\rm H\beta}$ measured along four MOSFIRE slits for both the restricted regions (shaded in blue slashes) and extended regions (shaded in green slashes) (see \S\ref{sec:spatial_oiii} for the definitions of restricted and extended regions).  
    \label{fig:Ratio}}
\end{figure}

Theoretically, we consider two principal scenarios of \lya\ production: (1) photo-ionization + recombination (e.g., due to star formation); (2) collisional excitation + radiative de-excitation (e.g., due to cold accretion). For scenario 1, assuming case B recombination, we use the \texttt{PyNeb} package \citep{Luridiana15} to calculate $F_{{\rm Ly}\alpha}$/$F_{\rm H\beta}$ for $T_{\rm \HI}$\,(K) $\in$ [10$^3$, 10$^5$] and $n_{\rm e}$\,(cm$^{-3}$) $\in$ [1, 10$^4$], where $T_{\rm \HI}$ and $n_{\rm e}$ are the kinetic temperature of the \HI\ gas and electron number density, respectively. For scenario 2, assuming collisional ionization equilibrium, we use the \texttt{ChiantiPy} package \citep{Dere1997, Dere13, Dere19} to calculate $F_{{\rm Ly}\alpha}$/$F_{\rm H\beta}$ for the same ranges of $T_{\rm \HI}$ and $n_{\rm e}$ as above. The derived $F_{{\rm Ly}\alpha}$/$F_{\rm H\beta}$ as a function of $T_{\rm \HI}$ for both scenarios are shown in Figure \ref{fig:Ratio}. It can be seen that as $T_{\rm \HI}$ increases, the predicted $F_{{\rm Ly}\alpha}$/$F_{\rm H\beta}$ is roughly constant for scenario 1, but decreases for scenario 2, as \lya\ emissivity drops more quickly than \hb.

Now we compare our spectroscopic data to the theoretical predictions above. For each MOSFIRE slit in Figure \ref{fig:Lya_images}, we construct a corresponding pseudo-slit to extract a 2D spectrum from the 3D KCWI datacube via a 3D datacube visualization tool \texttt{QFitsView} \citep{Davies10, Ott12View}. We then integrate the flux density in the wavelength dimension for each line and convert it to a SB accounting for the slit width. 

In Figure \ref{fig:OIII_Slits}, we show the line SB for \lya, \oiii\ and \hb\ along each slit. Evidently, \lya\ is not necessarily co-spatial with \oiii\ or H$\beta$, and is usually more extended along the slit. We further calculate $F_{\rm \oiii}$/$F_{\rm H\beta}$ (shown in red numbers) by integrating $SB_{\rm H\beta}$ and $SB_{\rm \oiii}$ along the slits for each identified source. We also calculate $F_{\rm Ly\alpha}$/$F_{\rm H\beta}$ in two ways, where $F_{\rm Ly\alpha}$ is calculated either by integrating $SB_{\rm Ly\alpha}$ over the same region as \hb\ and \oiii\ (the `restricted' region, as indicated by solid arrows), or by integrating over the full extent of \lya\ (the `extended' region, as indicated by dashed arrows). The results are shown next to the arrows (red for $F_{\rm \oiii}$/$F_{\rm H\beta}$ and green for $F_{\rm Ly\alpha}$/$F_{\rm H\beta}$).

\begin{figure*}

\centering
\includegraphics[width=0.9\linewidth]{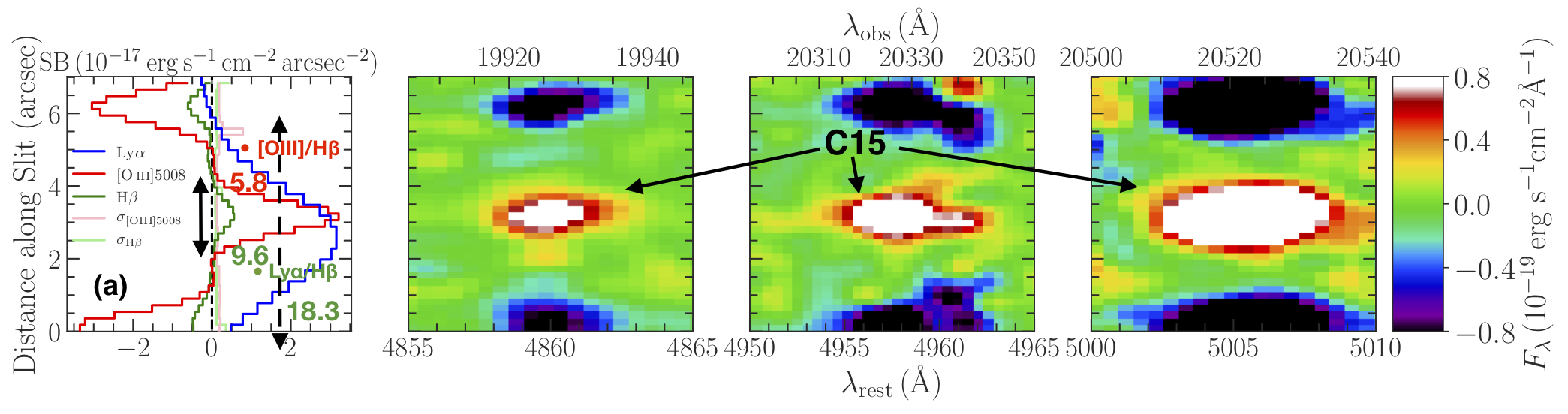}
\includegraphics[width=0.9\linewidth]{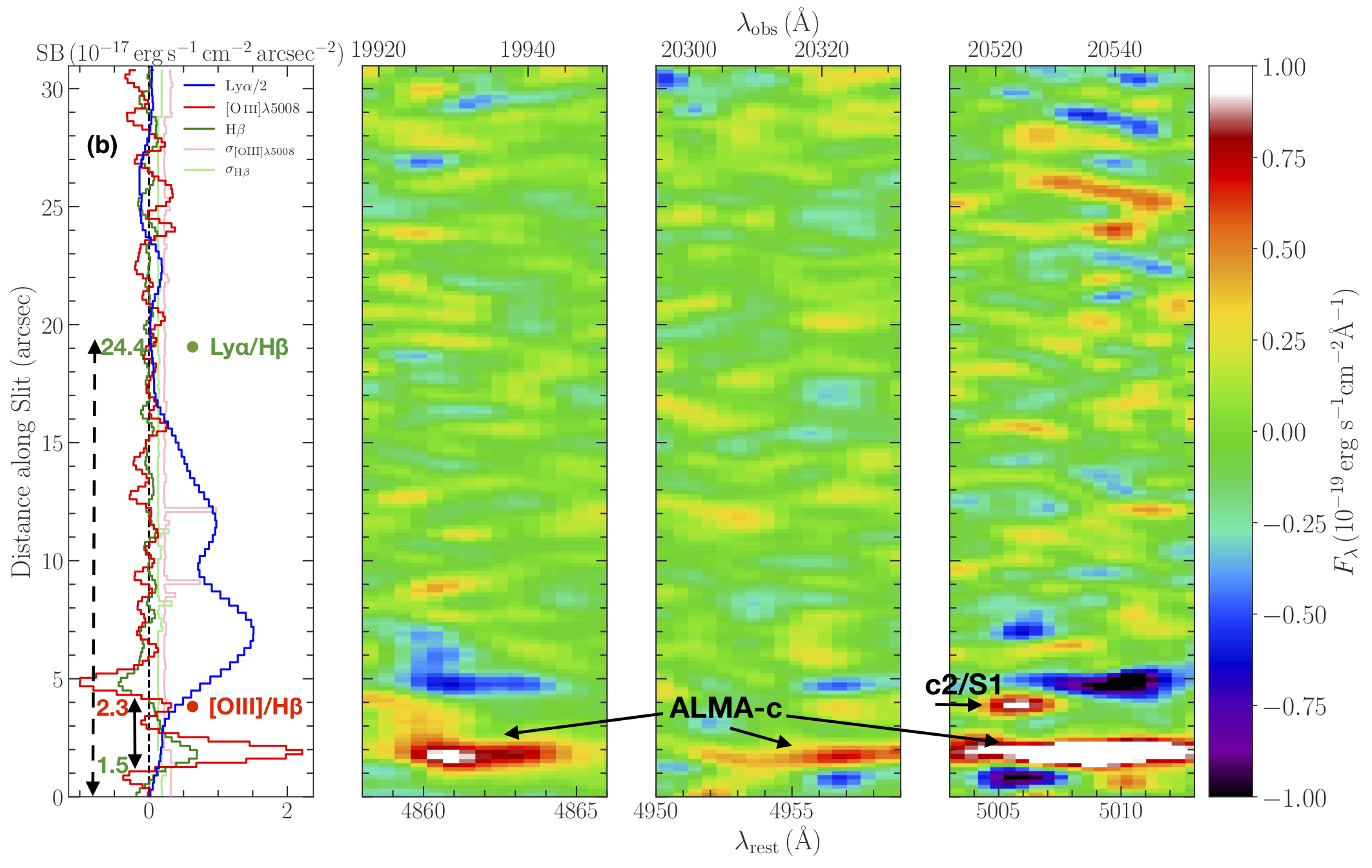}
\includegraphics[width=0.9\linewidth]{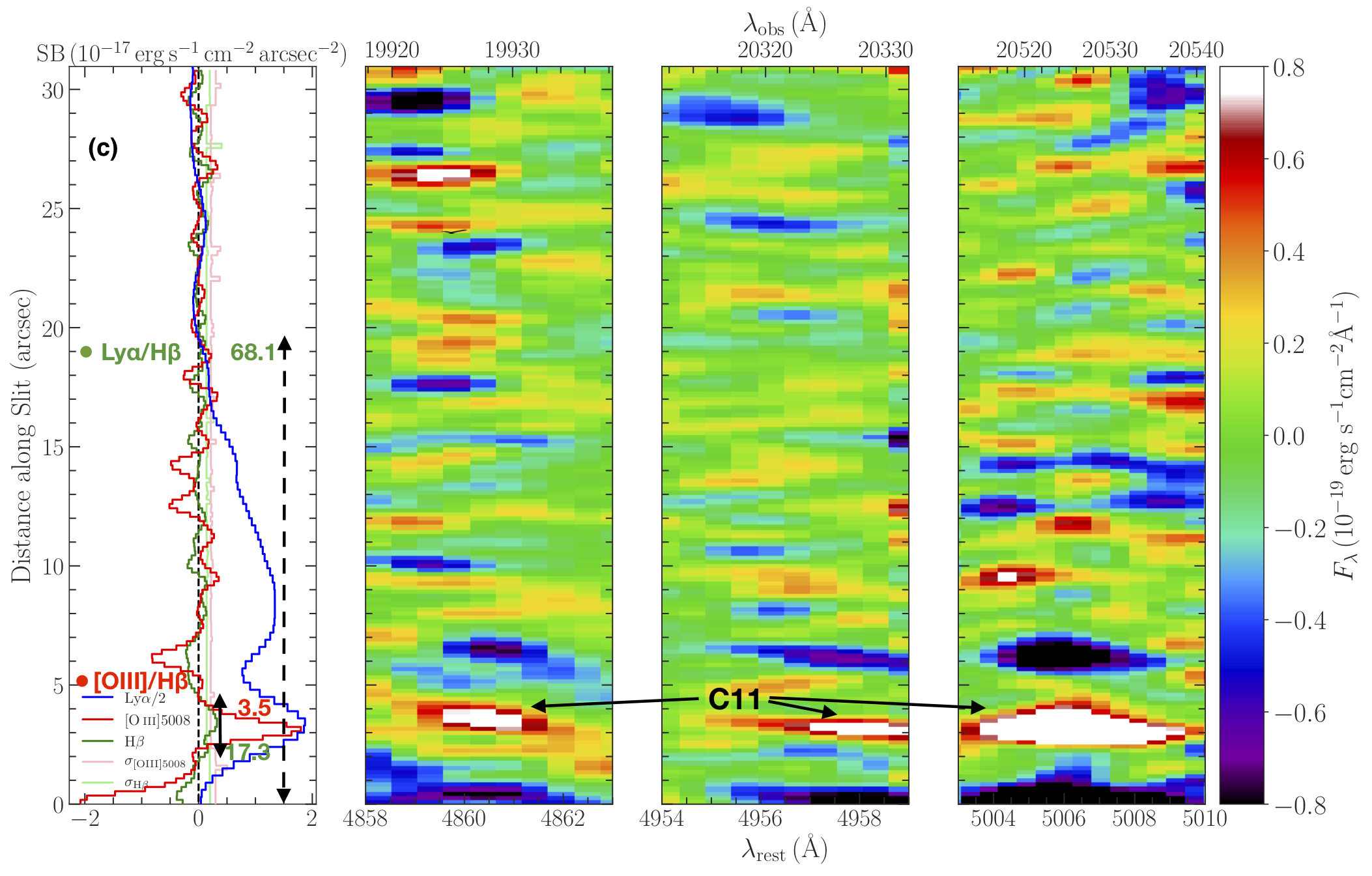}

\end{figure*}

\begin{figure*}

\centering
\includegraphics[width=0.89\linewidth]{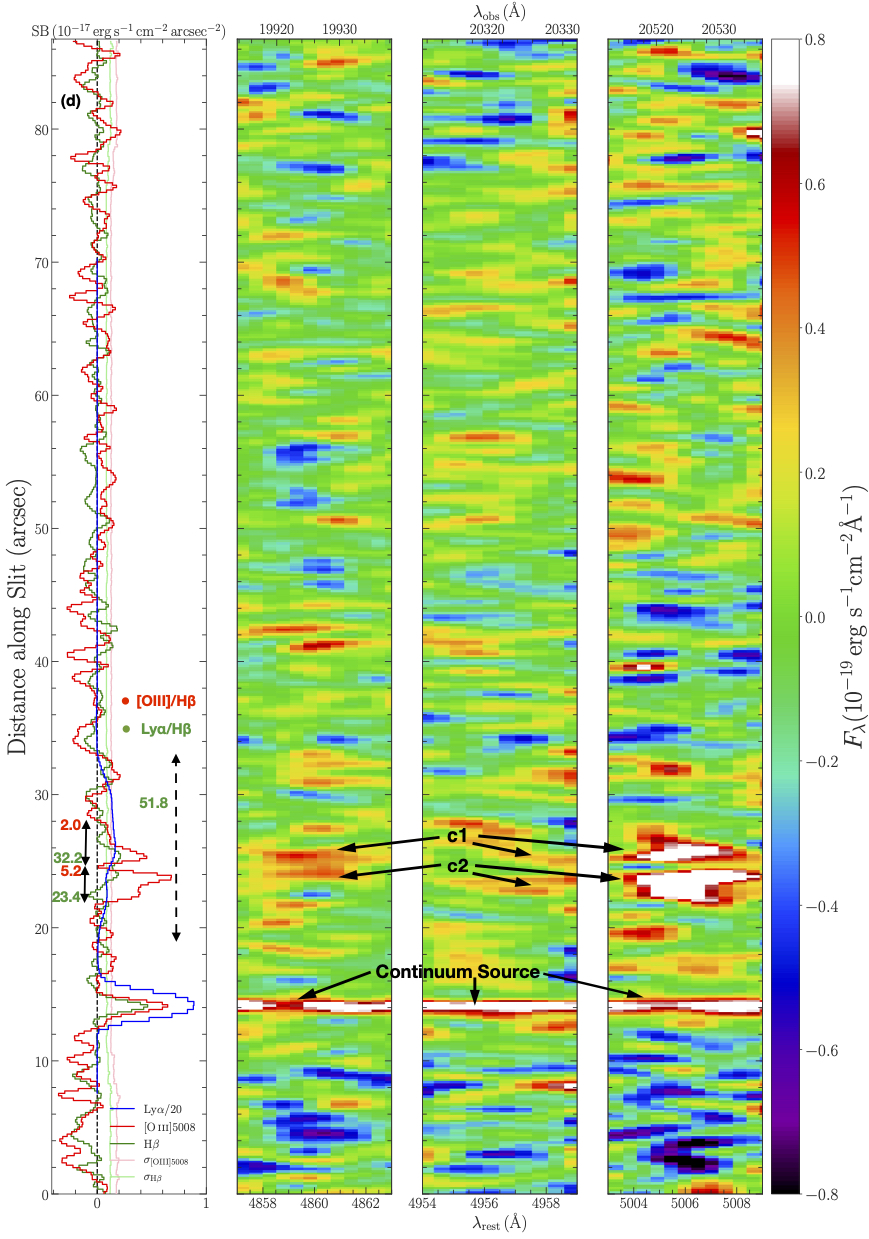}

\caption{Surface brightness distributions of \lya\ (blue), \oiii\ (red) and H$\beta$ (green) along four MOSFIRE slits (the leftmost panels, (a) - (d) correspond to slits 1 - 4, respectively). For reference, the smoothed 2D spectra of \hb\ (the second column) and \oiii\ (the third and fourth columns, for 4960\,\AA\ and 5008\,\AA, respectively) are also shown. Both the observed wavelengths ($\lambda_{\rm obs}$) and the rest-frame wavelengths ($\lambda_{\rm rest}$, assuming $z$ = 3.1000) are shown. The positions of known sources are indicated with black arrows. We calculate $F_{\rm \oiii}$/$F_{\rm H\beta}$ (shown in red numbers) by integrating $SB_{\rm H\beta}$ and $SB_{\rm \oiii}$ along the slits for each identified source. We also calculate $F_{\rm Ly\alpha}$/$F_{\rm H\beta}$ (shown in green numbers) in two ways, where $F_{\rm Ly\alpha}$ is calculated either by integrating $SB_{\rm Ly\alpha}$ over the same region as \hb\ and \oiii\ (as indicated by solid arrows), or by integrating over the full extent of \lya\ (as indicated by dashed arrows). Note that the blue regions in the 2D spectra are negative images due to dithering.}
\label{fig:OIII_Slits}
\end{figure*}

\begin{table*}
\renewcommand\arraystretch{1.5}
  \caption{\lya, \oiii\ and \hb\ fluxes and line ratios measured along four MOSFIRE slits.}
  \label{tab:line_fluxes}
  \centering
  \begin{tabular}{c|ccccc|cc}
    \hline \hline 
    \multicolumn{1}{c|}{}&\multicolumn{5}{c|}{Restricted Regions}&\multicolumn{2}{c}{Extended Regions}\\
    Slit No. & \lya & \oiii & \hb & \lya/\hb &\oiii/\hb &\lya &\lya/\hb \\
    (1)  & (2) & (3) & (4) & (5) & (6) & (7) & (8)\\
    \hline
    1  & 4.8 & 2.3 & 0.4 & 9.6 & 5.8 & 9.3 & 18.3\\
    2  & 1.3 & 1.5 & 0.7 & 1.5 & 2.3 & 20.9 & 24.4\\
    3  & 7.5 & 1.2 & 0.3 & 17.3 & 3.5 & 29.3 & 68.1\\
    4  & 9.2/4.3 & 0.6/1.1 & 0.3/0.2 & 32.2/23.4 & 2.0/5.2 & 24.3 & 51.8\\
    \hline \hline
  \end{tabular}
  \begin{tablenotes}
    \item \textbf{Notes.} \lya, \oiii\ and \hb\ fluxes measured by integrating along the four MOSFIRE slits (see Figure \ref{fig:Lya_images} and Table \ref{tab:mosfire_obs} for details) and the corresponding extracted pseudo-slits from the KCWI datacube and line ratios (\lya/\hb\ and \oiii/\hb). The columns are: (1) the slit number (as marked in Figure \ref{fig:Lya_images}); (2) the \lya\ flux of the restricted regions (where $F_{\rm Ly\alpha}$ is calculated over the same region as \hb\ and \oiii, as indicated by solid arrows in Figure \ref{fig:OIII_Slits}); (3) the \oiii\ flux; (4) the \hb\ flux; (5) the \lya/\hb\ ratio; (6) the \oiii/\hb\ ratio; (7)(8) same as (2)(5), but for extended regions (where $F_{\rm Ly\alpha}$ is calculated over the full extent of \lya, as indicated by dashed arrows in Figure \ref{fig:OIII_Slits}). All fluxes are in units of 10$^{-17}$\,erg\,s$^{-1}$\,cm$^{-2}$.
   \end{tablenotes}
\end{table*}

The results in Figure \ref{fig:Ratio} and \ref{fig:OIII_Slits} show that: (1) For slit 1 and 2, $F_{\rm Ly\alpha}$/$F_{\rm H\beta}$ are always smaller than the predicted value of scenario 1. Considering that \lya\ is subject to heavier dust extinction\footnote{Additionally, the scattering of \lya\ photons out of the line-of-sight can also reduce the observed $F_{\rm Ly\alpha}$/$F_{\rm H\beta}$ ratio.} than \hb, this result suggests that scenario 1 itself is sufficient to explain the observed $F_{\rm Ly\alpha}$/$F_{\rm H\beta}$; (2) For slit 3 and 4, we do see $F_{\rm Ly\alpha}$/$F_{\rm H\beta}$ ratios ($\sim$\,50 -- 70, see Table \ref{tab:line_fluxes}) higher than that predicted by scenario 1, but still far lower than those predicted by scenario 2 (especially in the $T_{\rm \HI}$ $\geq$ 10$^4$\,K region, where both \lya\ and \hb\ have been sufficiently excited). Simply scenario 1 and resonant scattering are sufficient to explain all the observed line ratios. Furthermore, we do not see a significant number of \lya\ profiles that have a blue dominant peak (signature of cold accretion, see e.g., \citealt{Zheng02, Dijkstra06b, Faucher10}) in these regions. Therefore, it is highly likely that photo-ionization + recombination is the main source of \lya\ photons, and resonant scattering (as indicated by significant polarization detections from \citealt{Hayes11} and \citealt{Beck16}) has substantially altered their spatial and kinematic distribution.

\subsection{Profiles of \lya\ Emission}\label{sec:profiles}

In this section, we investigate the variations of spatially-resolved \lya\ profiles in major emitting regions. Before proceeding, we first use \oiii\ to determine the systemic redshifts of three associated sources: C11, C15 and c$_1$. Single Gaussian fits to the \oiii\ line profiles (see Appendix \ref{sec:C11_C15} for details) yield redshifts (after heliocentric corrections) \emph{z}(C11) = 3.0980, \emph{z}(C15) = 3.0975\footnote{Compared to \citet{McLinden13}, our measurements for \emph{z}(C15) are consistent but our \emph{z}(C11) are slightly different. This may be due to: (1) the asymmetric nature of the \oiii\ profile of C11; (2) the misalignment between the MOSFIRE slit and the galaxy continuum emission.} and \emph{z}(c$_1$) = 3.0988, which we adopt as fiducial redshifts in the following analysis.

We then visualize the spatial variations of \lya\ peak position ($v_{\rm Ly\alpha}$) and line width ($\sigma_{\rm Ly\alpha}$) by making moment maps. The first and second flux-weighted moments are defined as:

\begin{align}\label{eq:mu1}
v_{{\rm Ly\alpha}, xy} = \frac{ \sum_{k}{v_{xyk} I_{xyk} } }{\sum_{k}{I_{xyk} } }
\end{align}

\begin{align}\label{eq:mu2}
\sigma_{{\rm Ly\alpha}, xy} =  \sqrt{ \frac{ \sum_{k}{ (v_{xyk}-v_{{\rm Ly\alpha}, xy})^2 I_{xyk} } }{\sum_{k}{I_{xyk} } }}
\end{align}
\noindent
where $I_{xyk}$ and $v_{xyk}$ are the flux density and velocity (relative to a fiducial redshift) of the $k$th wavelength layer at position $(x, y)$. In our moment analysis we fix the fiducial redshift of LAB1 at $z$ = 3.1, and all the summations are carried out over 4959 -- 5009\,\AA. Before applying Eqs. (\ref{eq:mu1}) and (\ref{eq:mu2}), we filter out all the voxels with S/N < 6 (for $v_{\rm Ly\alpha}$) or 4 (for $\sigma_{\rm Ly\alpha}$)\footnote{Our experiments show that these choices maximize the inclusion of real signal without introducing spurious detections.}. The $\sigma_{\rm Ly\alpha}$ map has been further corrected for the KCWI instrumental line spread function (LSF), $\sigma$ = 65\,km\,s$^{-1}$.

The resulting moment maps are shown in Figure \ref{fig:Lya_kinematics}. The two major \lya\ emitting regions have been delineated by rectangular boxes. We zoom in on these two regions in separate panels. By adjusting the dynamic range, we are able to discern more subtle structures, discussed in the following sections (\S\ref{sec:upper}-\ref{sec:lower}).

\subsubsection{Northern Region}\label{sec:upper}
There is only one identified source (C15) in the northern \lya\ emitting region. We first use a large aperture to measure the global properties of LBG C15. The global line widths of \lya\ and \oiii\ of C15 are 250 and 64\,km\,s$^{-1}$ (corrected for LSF, $\sigma$ = 65\,km\,s$^{-1}$ for KCWI and 35\,km\,s$^{-1}$ for MOSFIRE). The global velocity offset between \lya\ and \oiii\ is $\Delta v_{\rm Ly\alpha}$ = --22\,km\,s$^{-1}$, although it varies at different locations. This $\Delta v_{\rm Ly\alpha}$ is significantly smaller than the velocity offsets observed in LBGs \citep{Steidel10} and LAEs \citep{McLinden11}, both of which are $\gtrsim$\,300\,km\,s$^{-1}$ and are interpreted as signs of outflows. Therefore, it is tempting to conclude that this region should be lack of significant outflows. However, as we will show in Section \ref{sec:results}, our multiphase, clumpy model predicts that significant outflow velocities can still be present in profiles with $\Delta v_{\rm Ly\alpha} \simeq$ 0 (e.g., in our spectra 1 and 2).

\begin{figure*}
\includegraphics[width=0.91\textwidth]{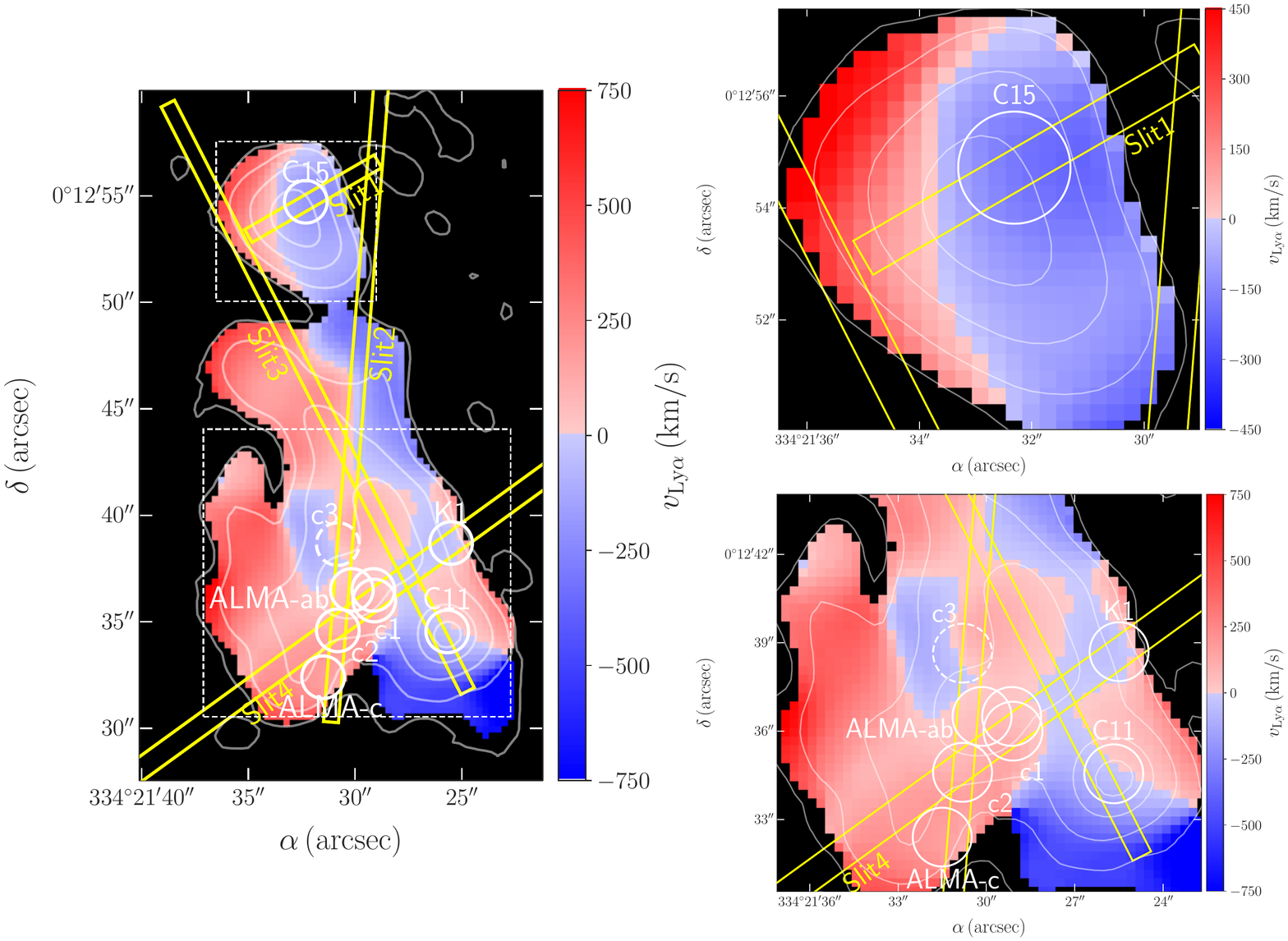}\\
\includegraphics[width=0.91\textwidth]{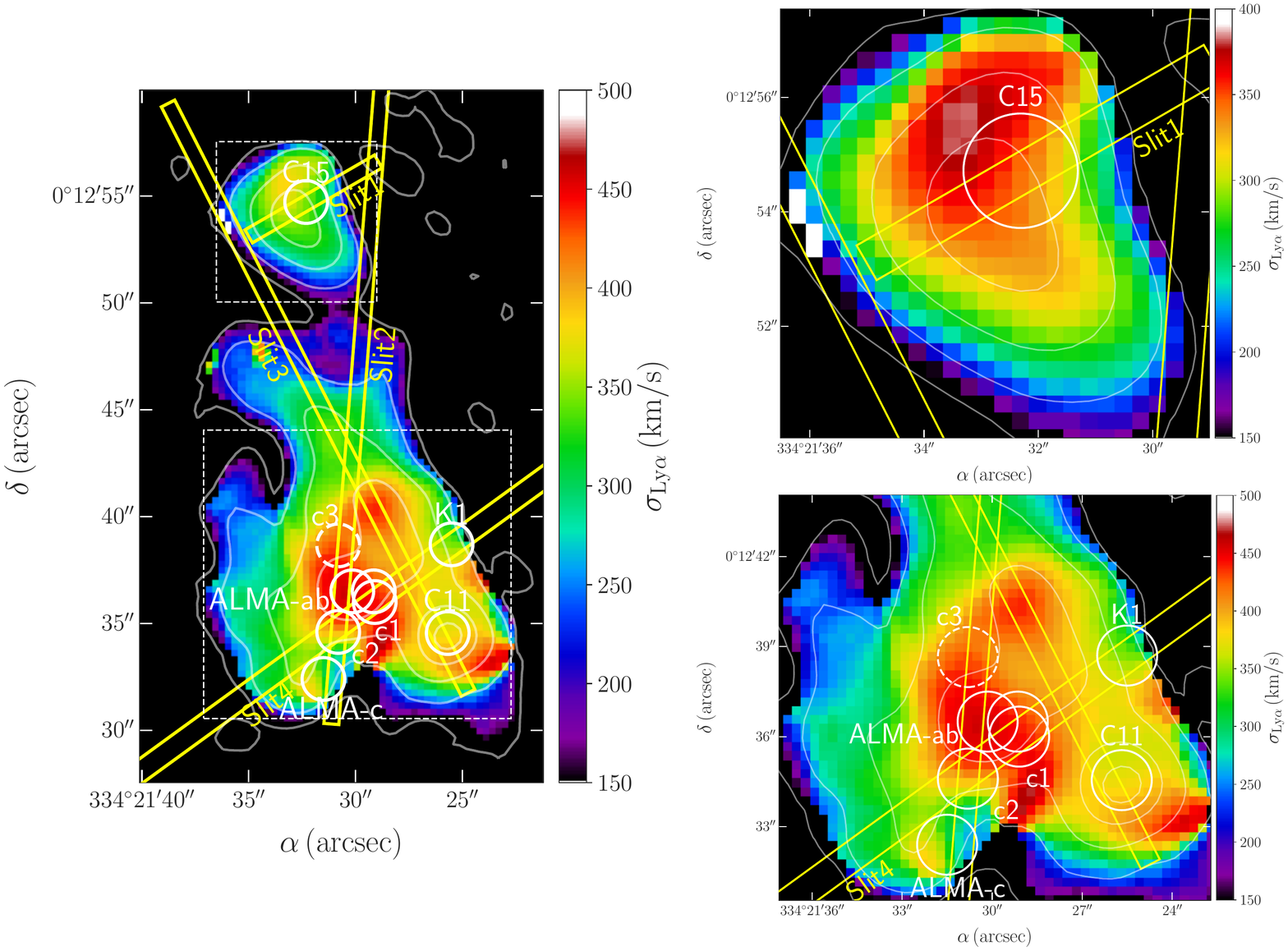}\\
    \caption{The first ($v_{\rm Ly\alpha}$) and second ($\sigma_{\rm Ly\alpha}$) moment maps of LAB1. The two major \lya\ emitting regions have been delineated by rectangular boxes (dashed white lines), and their zooming-in views are shown in the right panels. For the northern region, the colorbar limits have been adjusted accordingly to account for the smaller value range. The positions of the identified continuum sources are indicated by circles with labels. The \lya\ SB isophotes (solid white lines) with levels of SB$_{\rm Ly\alpha}$ = [120,\,80,\,40,\,15,\,4]$\,\times\,10^{-19}$\,erg\,s$^{-1}$\,cm$^{-2}$\,arcsec$^{-2}$ have also been overlaid onto each image for visual reference.  
    \label{fig:Lya_kinematics}}
\end{figure*}

Most \lya\ profiles in the northern region are considerably asymmetric and consist of a `main peak' and a `red bump' (see spectrum 1 in Figure \ref{fig:Lya_linemaps} as an example). Moreover, the main peak is redshifted towards the eastern region, and blueshifted towards the west. The largest $v_{\rm Ly\alpha}$ can be up to $\sim$\,500\,km\,s$^{-1}$, which explains the evident east-west $v_{\rm Ly\alpha}$ gradient in Figure \ref{fig:Lya_kinematics}. This shear in $v_{\rm Ly\alpha}$ appears to be perpendicular to the major axis of C15, which is consistent with the suggestion by \citet{Weijmans10} that outflow or rotation is indicated. 

As for $\sigma_{\rm Ly\alpha}$, its largest value ($\sim$\,400\,km\,s$^{-1}$) is located slightly north-east of C15, beyond which $\sigma_{\rm Ly\alpha}$ gradually decreases moving away from C15. In general, the $\sigma_{\rm Ly\alpha}$ values in the northern region are much larger than the global $\sigma_{\rm \oiii}$. This is unexpected if one were to assume that both \lya\ and \oiii\ photons are emitted by the same sources, unless the kinematics of \lya\ have been altered by radiative transfer effects. We attempt to explain the broadening of \lya\ in Section \ref{sec:RT}. 

\subsubsection{Southern Region}\label{sec:lower}
Multiple discrete continuum sources have been identified within the southern portion of LAB1, including the LBG C11, three ALMA submm sources, and several very faint objects with spectroscopic confirmation (K1, c$_{1}$, c$_{2}$ and c$_{3}$). We first use a large aperture to measure the global properties of LBG C11. The LSF-corrected global line widths of \lya\ and \oiii\ of C11 are 178 and 78\,km\,s$^{-1}$. The global velocity offset between \lya\ and \oiii\ is $\Delta v_{\rm Ly\alpha}$ = +175\,km\,s$^{-1}$ (i.e., redshifted with respect to systemic), and $-197$\,km\,s$^{-1}$ between the Si ${\textsc {ii}}$ 1526 absorption line (from LRIS observations) and \lya. Similar velocity offsets between \lya\ and interstellar absorption features are commonly observed in `down the barrel' spectra of LBGs, and are generally interpreted as signatures of outflow \citep{Steidel10}. However, they are inconsistent with the non-detection of s significant offset between \lya\ and \oiii\ by \citet{McLinden11}. This may be due to the high asymmetry of the \oiii\ profile of C11.

\begin{figure*}
\centering
\includegraphics[width=\textwidth]{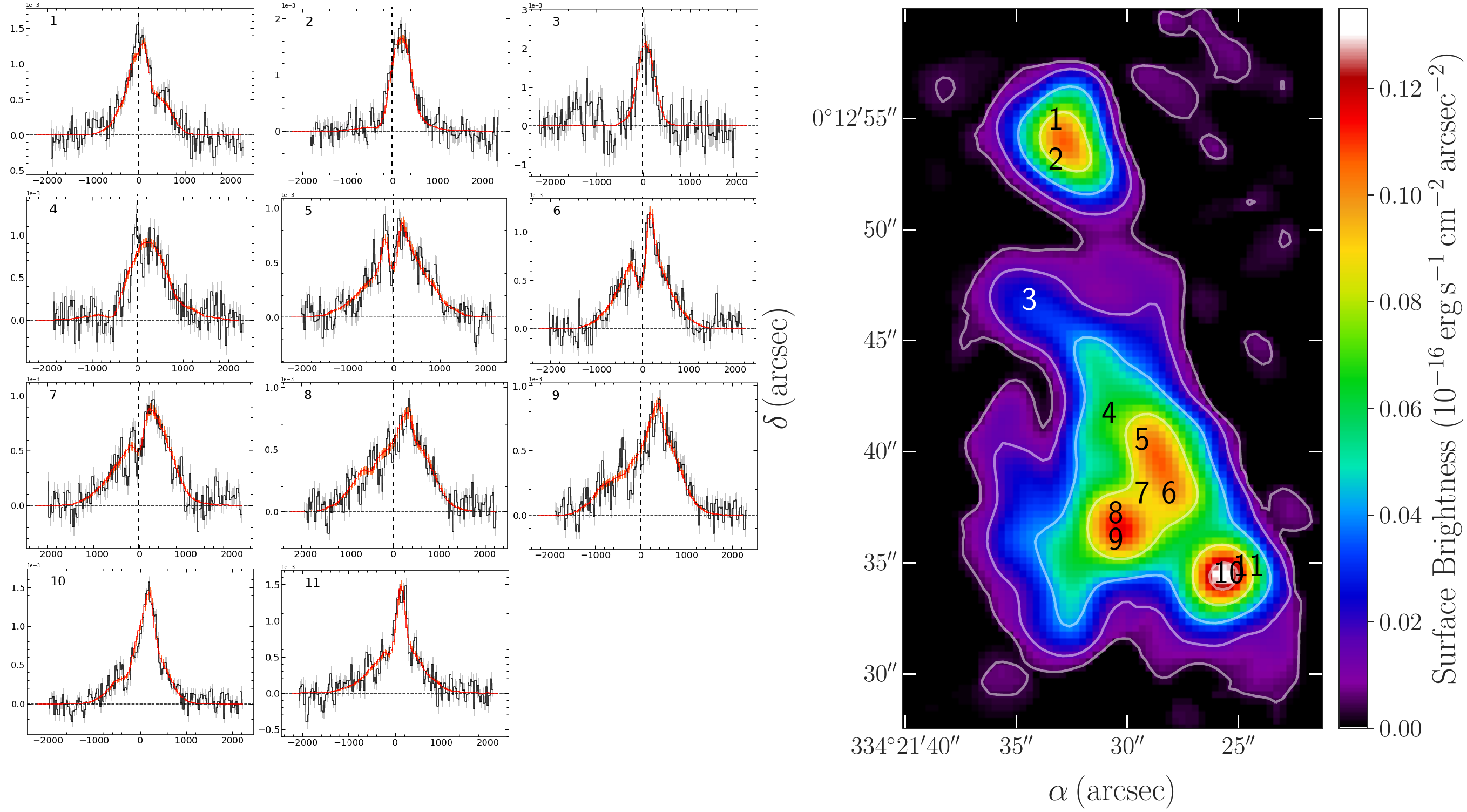}
    \caption{Eleven representative continuum-subtracted, spatially-resolved \lya\ profiles from the high SB regions in LAB1. All the spectra have been smoothed by a 3 pixel $\times$ 3 pixel boxcar (0.9$\arcsec$) spatially and Gaussian smoothed ($\sigma$ = 0.5\,\AA) in the wavelength dimension. The multiphase clumpy model best-fits (red, with orange 1-$\sigma$ Poisson errors) and the observed \lya\ profiles (black, with grey 1-$\sigma$ error bars) have both been normalized. The observed \lya\ spectra have also been shifted by --$\Delta v$ to their local systemic redshifts. For each subpanel, the $x$-axis is the velocity (in km\,s$^{-1}$) with respect to the local systemic redshift, and the $y$-axis is the normalized line flux. The spectrum number of each spectrum has been marked on the SB map (the right panel). For visual reference, the horizontal and vertical black dashed lines in each subpanel indicate zero flux level and zero velocity with respect to the local systemic redshift, respectively. 
    \label{fig:Lya_linemaps}}
\end{figure*}

Most \lya\ profiles from spatial locations near C11 exhibit double peaks -- a red dominant peak + a blue `bump' (see spectrum 10 in Figure \ref{fig:Lya_linemaps} as an example). The position of the red dominant peak tends to move towards more blueshifted velocities along the northwest-southeast direction. The largest $v_{\rm Ly\alpha}$ is $\sim$\,300\,km\,s$^{-1}$, which gives rise to the $v_{\rm Ly\alpha}$ northwest-southeast gradient in Figure \ref{fig:Lya_kinematics}. This shear in $v_{\rm Ly\alpha}$ appears to be parallel to the major axis of C11, consistent with \citet{Weijmans10}.

As for $\sigma_{\rm Ly\alpha}$, its largest value ($\sim$\,500\,km\,s$^{-1}$) is located in the southwest corner, while the majority of the spectra around C11 have a rather homogeneous $\sigma_{\rm Ly\alpha}$\,$\sim$\,400\,km\,s$^{-1}$. Again, these values are much larger than $\sigma_{\rm \oiii}$. 

The \lya\ profiles near the ALMA sources are more complex -- most of them are very broad, highly asymmetric, and have multiple peaks. Some of the profiles (e.g., the northeast corner) are even dominated by a `blue peak', as shown in spectrum 4 in Figure \ref{fig:Lya_linemaps}. 

On the $v_{\rm Ly\alpha}$ map, there is an alternate pattern of positive and negative $v_{\rm Ly\alpha}$ from the east to the west. Yet again, we see a similar coherent velocity structure that coincides with the high SB `tadpole' structure (see Section \ref{sec:spatial}). This structure is also seen on the $\sigma_{\rm Ly\alpha}$ map, but with a slightly different trend -- starting from the south, first going towards northeast, and then turning northwest. The largest $\sigma_{\rm Ly\alpha}$ values ($\sim$\,500\,km\,s$^{-1}$) still overlap with ALMA-a, which indicates that the ALMA source may be responsible for the \lya\ line broadening (e.g., via starburst-driven outflows).

Our $v_{\rm Ly\alpha}$ and $\sigma_{\rm Ly\alpha}$ maps are qualitatively similar to the ones presented in a recent work by \citet{Herenz20}, albeit with slight differences in the extent of the \lya\ emitting regions and the number of spaxels included, due to different FOVs of instruments and S/N threshold choices. The alternate pattern of positive and negative $v_{\rm Ly\alpha}$ is consistent with the left panel in their Figure 7, and the large $\sigma_{\rm Ly\alpha}$ values near ALMA-ab sources are consistent with the right panel in their Figure 7.
 
\section{Radiative Transfer Modeling Using The Multiphase Clumpy Model}\label{sec:RT}

\begin{table*}
\renewcommand\arraystretch{1.8}
  \scriptsize \caption{Fitted parameters of the multiphase clumpy model and derived quantities.}
  \label{tab:Fitting_results}
  \centering
  \begin{tabular}{ccc|cccccc|cc|cc}
    \hline \hline 
    \multicolumn{3}{c|}{}&\multicolumn{6}{c|}{Fitted Parameters}&\multicolumn{2}{c|}{Derived Parameters}&\multicolumn{2}{c}{Moments}\\
    No. & RA (J2000) & Dec (J2000) & ${\rm log}\,n_{\rm HI,\,{\rm ICM}}$ & $F_{\rm V}$ & $\sigma_{\rm cl}$ & $v_{\rm cl}$ & $v_{\rm ICM}$ & $\Delta v$ & $f_{\rm cl}/f_{\rm cl, crit}$ & $\log\tau_{\rm 0, ICM}$ & $v_{\rm Ly\alpha}$ & $\sigma_{\rm Ly\alpha}$\\
    & & &(cm$^{-3}$) & &(km\,s$^{-1}$)   & (km\,s$^{-1}$) & (km\,s$^{-1}$) &(km\,s$^{-1}$) & & &(km\,s$^{-1}$)&(km\,s$^{-1}$)\\
    (1) & (2) & (3) & (4) & (5) & (6) & (7) & (8) & (9) & (10) & (11) & (12) & (13) \\
    \hline
    1 & 22:17:26.214 & +00:12:54.85 & --7.13$^{+0.10}_{-0.13}$ &  0.37$^{+0.15}_{-0.16}$ & 131$^{+167}_{-94}$ & 773$^{+26}_{-159}$ &8$^{+19}_{-7}$ & --195$^{+34}_{-49}$ & $12.8^{+17.2}_{-10.2}$ & $0.8^{+0.1}_{-0.1}$  & --98.2 & 373.9 \\
    2 & 22:17:26.214 & +00:12:53.05 & --6.47$^{+0.40}_{-0.80}$ & 0.37$^{+0.22}_{-0.25}$ & 242$^{+112}_{-104}$ & 497$^{+217}_{-203}$ &444$^{+131}_{-60}$ & --268$^{+60}_{-88}$ & $0.8^{+1.8}_{-0.7}$  &  --3.6$^{+1.0}_{-0.8}$  & --54.2 & 324.2 \\
    3 & 22:17:26.294 & +00:12:46.75 & --7.01$^{+0.88}_{-0.88}$ &  0.32$^{+0.26}_{-0.21}$ & 111$^{+263}_{-102}$ & 392$^{+385}_{-259}$ &348$^{+156}_{-96}$ &   43$^{+92}_{-129}$& $0.1^{+6.9}_{-0.1}$ & --2.1$^{+1.5}_{-2.8}$  & 193.4 & 254.1 \\
    4 & 22:17:26.054 & +00:12:41.65 & --6.15$^{+0.14}_{-0.31}$ & 0.18$^{+0.25}_{-0.07}$ & 705$^{+91}_{-392}$ & 689$^{+108}_{-494}$ &653$^{+123}_{-62}$ & --200$^{+48}_{-90}$& $0.7^{+0.7}_{-0.6}$ & --4.2$^{+0.2}_{-0.3}$ & 20.6  & 339.2 \\
    5 & 22:17:25.954 & +00:12:40.45 & --6.94$^{+0.09}_{-0.57}$ &  0.23$^{+0.09}_{-0.03}$ & 534$^{+53}_{-91}$ & 474$^{+205}_{-463}$ &7$^{+58}_{-7}$ &   --69$^{+139}_{-27}$& $2.0^{+1.4}_{-0.4}$ &  1.0$^{+0.1}_{-0.7}$  & 81.5 & 435.2 \\
    6 & 22:17:25.874 & +00:12:38.05 & --6.90$^{+0.10}_{-0.26}$& 0.18$^{+0.04}_{-0.02}$ & 455$^{+48}_{-47}$ & 208$^{+234}_{-196}$ &   48$^{+46}_{-28}$ &--66$^{+58}_{-32}$& $1.8^{+0.5}_{-0.3}$ &  1.0$^{+0.1}_{-0.4}$  & 6.2 & 394.9 \\
    7 & 22:17:25.954 & +00:12:38.05 & --7.02$^{+0.14}_{-0.90}$ &  0.29$^{+0.19}_{-0.06}$ & 437$^{+48}_{-41}$ & 376$^{+127}_{-188}$ &16$^{+391}_{-15}$ &   --100$^{+53}_{-44}$& $3.0^{+1.6}_{-0.8}$ &  0.9$^{+0.1}_{-5.1}$  & 46.3 & 402.5 \\
    8 & 22:17:26.034 & +00:12:37.15 & --7.90$^{+0.53}_{-0.10}$ & 0.51$^{+0.04}_{-0.05}$ & 498$^{+77}_{-39}$ & 228$^{+98}_{-76}$ &435$^{+138}_{-112}$ & --162$^{+61}_{-65}$& $4.3^{+0.5}_{-0.8}$  &  --4.8$^{+2.2}_{-1.0}$  & 14.6 & 448.8 \\
    9 & 22:17:26.034 & +00:12:35.95 & --7.83$^{+0.72}_{-0.16}$ &  0.56$^{+0.03}_{-0.07}$ & 475$^{+47}_{-31}$ & 276$^{+68}_{-123}$ &425$^{+108}_{-108}$ & --181$^{+89}_{-62}$& $4.6^{+0.5}_{-0.6}$ &  --4.6$^{+2.5}_{-1.1}$  & 54.3 & 452.9 \\
    10 & 22:17:25.694 & +00:12:34.45 & --7.49$^{+0.33}_{-0.48}$ & 0.28$^{+0.15}_{-0.11}$ & 353$^{+110}_{-32}$ & 270$^{+75}_{-246}$ &302$^{+91}_{-70}$ &   --135$^{+109}_{-73}$& $2.8^{+1.1}_{-1.0}$ & --1.9$^{+1.1}_{-2.0}$ & -19.6 & 377.1 \\
    11 & 22:17:25.634& +00:12:34.75 & --7.32$^{+0.41}_{-0.64}$ &  0.19$^{+0.33}_{-0.05}$ & 453$^{+95}_{-111}$ & 268$^{+309}_{-256}$ &151$^{+260}_{-95}$ &   --9$^{+69}_{-180}$& $1.9^{+2.6}_{-0.7}$ &  0.0$^{+0.9}_{-4.3}$ & 44.7 & 378.4 \\
    \hline \hline
  \end{tabular}
  \begin{tablenotes}
  \small
    \item \textbf{Notes.} Fitted parameters (averages and 2.5\% -- 97.5\% quantiles, i.e., 2-$\sigma$ confidence intervals) of the multiphase clumpy model, derived quantities and spectral moments. The columns are: (1) the spectrum number (as marked in Figure \ref{fig:Lya_linemaps}); (2) the right ascension of the center of the extracted region; (3) the declination of the center of the extracted region; (4) the \HI\ number density in the ICM; (5) the cloud volume filling factor; (6) the velocity dispersion of the clumps; (7) the radial outflow velocity of the clumps; (8) the \HI\ outflow velocity in the ICM; (9) the velocity shift relative to $z$ = 3.1000 (a negative/positive value means that the model spectrum has been blue/redshifted to match the data); (10) the clump covering fraction (defined as the number of clumps per line-of-sight) normalized by the critical clump covering fraction. In our case $f_{\rm cl}$ = 75\,$F_{\rm V}$. The critical clump covering fraction, $f_{\rm cl, crit}$, determines different physical regimes and is calculated via Eq. (\ref{eq:fccrit}) (see Appendix \ref{sec:fccrit} for a detailed derivation); (11) the optical depth at the \lya\ line center of the \HI\ in the ICM; (12) the first moment of the center of the extracted region; (13) the second moment of the center of the extracted region (corrected for KCWI LSF). 
   \end{tablenotes}
\end{table*}

\begin{figure*}
\centering
\includegraphics[width=\textwidth]{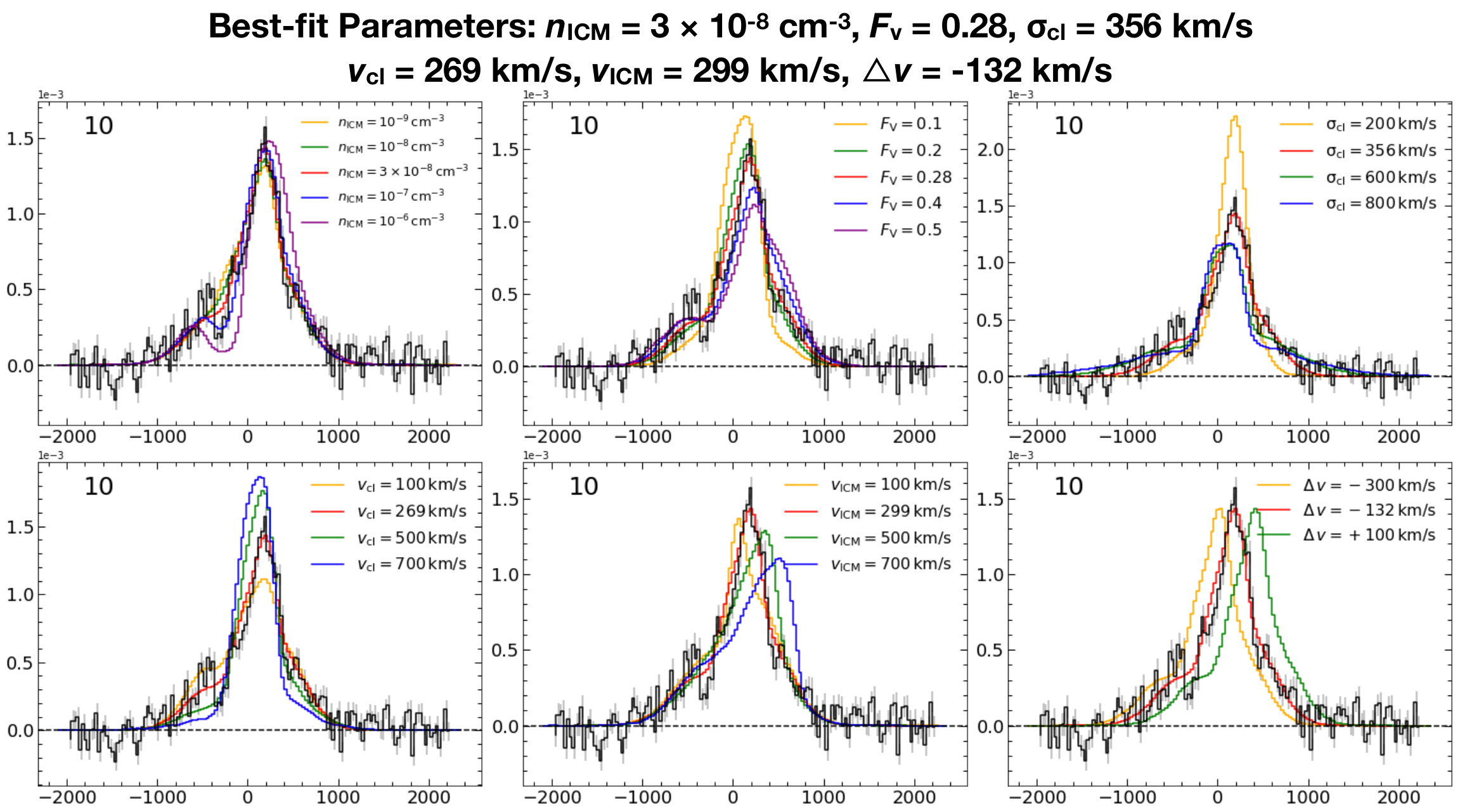}
    \caption{The effects of each individual physical parameter in [$n_{\rm HI,\,{\rm ICM}}, F_{\rm V}, \sigma_{\rm cl}, v_{\rm cl}, v_{\rm ICM}, \Delta v$] (taking spectrum 10 as an example). From the top left to the bottom right panel, one parameter is varied at a time (as shown in lines and labeled with different colors) and others are fixed (to the best-fit parameter values of spectrum 10, see Table \ref{tab:Fitting_results}). The red line in each panel represents the best-fit model of spectrum 10. The $x$-axis is the velocity (in km\,s$^{-1}$) with respect to the local systemic redshift of spectrum 10, and the $y$-axis is the normalized line flux. It can be seen that different parameters affect the model spectra in different ways -- $n_{\rm HI,\,{\rm ICM}}$ determines the overall shape and the trough depth of the spectrum; $F_{\rm V}$ and $v_{\rm cl}$ determine the shapes and strengths of the peak(s); $\sigma_{\rm cl}$ determines the width of the spectrum; $v_{\rm ICM}$ determines the location of the main peak and the trough; $\Delta v$ shifts the spectrum in the velocity dimension.
    \label{fig:vary}}
\end{figure*}

\subsection{Methodology}

Although the moment map analysis above provides a cursory overview of the apparent gas velocity field, it is purely phenomenological and could even be misleading, in the sense that if radiative transfer effects dominate, the observed $v_{\rm Ly\alpha}$ would not necessarily be linked directly to the local gas kinematics. To gain more physical insight and to account for the possibly important radiative transfer effects, we generated a series of model spectra using MCRT and fit them to the observed \lya\ spectra at different positions in LAB1.

Our first attempt was to fit the \lya\ profiles using the widely used `shell model' \citep{Verhamme06,Dijkstra06b}. However, as most line profiles are fairly broad and multi-peaked with significant flux close to line center, the fits either fail to reproduce the major features or have inexplicably large intrinsic line widths (see e.g., \citealt{Orlitova18}). Therefore, we adopt a more sophisticated and physically realistic multiphase `clumpy model' instead. As described in \citet{Gronke16_model}, the geometric setup of this `clumpy model' is a number of spherical \HI\ clumps moving within a hot ($T\sim 10^{5-7}$\,K), ionized inter-clump medium (ICM) \citep[see also][]{Laursen13}. This model predicts the \lya\ spectra produced by a central \lya\ emitting source, accounting for the scattering by \HI\ (both in the clumps and the ICM)\footnote{As we will show below, the scattering process washes out the information about the \lya\ emitting source, i.e., the initial spatial or spectral shape of the source does not significantly affect the emergent spectra.}. It has 14 parameters in total (see the detailed formulation in \citealt{Gronke16_model}), among which the most important ones are the cloud covering factor ($f_{\rm {\rm cl}}$) that describes the mean number of clumps per line-of-sight from the center to the boundary of the simulation sphere, the \HI\ number density in the ICM ($n_{\rm HI,\,{\rm ICM}}$), and kinematic parameters of the clumps and ICM. Specifically, the clump motion is assumed to be a superposition of an isotropic Gaussian random motion (characterized by $\sigma_{\rm {\rm cl}}$, the velocity dispersion of the clumps) and a radial uniform outflow with a constant velocity $v_{\rm cl}$. In addition, we consider an outflow velocity of the low density \HI\ in the ICM, $v_{\rm ICM}$, and a post-processed parameter, the velocity shift with respect to $z$ = 3.1000, $\Delta v$. This $\Delta v$ parameter represents the best-fit systemic redshift of the \lya\ source function relative to $z$ = 3.1000 (the initial guess for the systemic redshift). 

Note that \lya\ radiative transfer in such a multiphase medium exhibits two characteristic regimes defined by the values of $f_{\rm cl}$. If $f_{\rm cl}$ is (much) greater than a critical value $f_{\rm cl, crit}$ (which is a function of other model parameters, such as the kinematics and \HI\ column density of the clumps), the photons would escape as if the medium is homogeneous and the emergent spectra are similar to the ones predicted by the aforementioned `shell  model' \citep{Gronke17b}. Otherwise, for fewer clumps per line-of-sight, the photons preferentially travel in the ionized ICM and escape closer to the line center of \lya. As most of our observed \lya\ spectra have considerable flux near the line center, we expect $f_{\rm cl}\lesssim f_{\rm cl, crit}$ in our cases, and will focus on that regime\footnote{Note that the model spectra are sensitive to $\sim\,f_{\rm cl}/f_{\rm cl, crit}$ but not to certain individual model parameters, such as the \HI\ column density or the shape of the clumps \citep{Hansen06}.}.

Based on these considerations, we further construct a five-dimensional hypercubic grid by varying five crucial physical parameters: [${\rm log}\,n_{\rm HI,\,{\rm ICM}}, F_{\rm V}, \sigma_{\rm cl}, v_{\rm cl}, v_{\rm ICM}$]\footnote{For convenience we vary $F_{\rm V}$ rather than $f_{\rm cl}$ when generating clumps, but they are directly related via $f_{\rm cl}$ = 3$r_{\rm gal}$/4$r_{\rm cl}$\,$F_{\rm V}$, where $r_{\rm gal}$ = 5\,kpc is the radius of the simulation sphere and $r_{\rm cl}$ = 50\,pc is the clump radius (hence $f_{\rm cl}$ = 75\,$F_{\rm V}$ in our case).}. The prior ranges of ${\rm log}\,n_{\rm HI,\,{\rm ICM}}$\,(cm$^{-3}$), $F_{\rm V}$ and [$\sigma_{\rm cl}, v_{\rm cl}, v_{\rm ICM}$]\,(km\,s$^{-1}$) are [--8, --6], [0.1, 0.6] and [0, 800] (with spacings of 0.4, 0.1 and 100), respectively. We fix the subdominant parameters, such as the ICM temperature $T_{\rm ICM}$ to 10$^{6}$\,K, and the clump column density to 10$^{17}$\,cm$^{-2}$ in order to reduce the dimensionality of the parameter space\footnote{Here we modeled each observed spectrum with a set of parameters of the scattering medium independently, whereas in reality the \lya\ photons are likely to be scattered by a common medium with spatially-varying parameters. The application of a more advanced, self-consistent model is beyond the scope of this work but will be explored in the future.}.

Such a configuration amounts to $26244$ models in total. Each model is calculated via radiative transfer using $10000$ \lya\ photon packages generated from a Gaussian intrinsic spectrum \emph{N}(0, $\sigma^2_{\rm cl}$), where $\sigma_{\rm cl}$ = 12.85\,km\,s$^{-1}$ is the canonical thermal velocity dispersion of $T$ = 10$^4$\,K gas\footnote{Note that we did not employ the commonly used `core-skipping' technique, as it may cause artifacts in a multiphase medium.}. The sixth parameter, $\Delta v$, is varied continuously in post-processing. To properly explore the multimodal posterior of the parameters, we further use a python nested sampling package \texttt{dynesty} \citep{Skilling04, Skilling06, Speagle20} to fit the \lya\ spectra. 

To demonstrate the feasibility of this fitting routine using clumpy models, we selected eleven representative \lya\ spectra from the high SB regions, as shown in Figure \ref{fig:Lya_linemaps}. All of the spectra presented have been smoothed spatially by a 3 pixel $\times$ 3 pixel boxcar and were extracted by using an $R$ = 3 pixel (0.9$\arcsec$) aperture. For each spectrum, we used 1500 live points (the initial randomly drawn samples from the prior) to calculate a set of clumpy models via linear flux interpolation on the grid and convolved with the KCWI LSF before comparing them to the observed \lya\ profiles. The best-fit spectra are also shown in Figure \ref{fig:Lya_linemaps}.

\subsection{Results}\label{sec:results}

In Figure \ref{fig:Lya_linemaps}, one can see that most of the model fits match the observations reasonably well. The values of the fitted parameters are presented in Table \ref{tab:Fitting_results}, and the derived joint and marginal posterior probability distributions are presented in Appendix \ref{sec:posterior}. We find that different parameters affect the model spectra in different ways -- $n_{\rm HI,\,{\rm ICM}}$ determines the overall shape of the spectrum and the depth of the intensity minimum near the systemic velocity (the `trough'); $F_{\rm V}$ and $v_{\rm cl}$ determine the shapes and strengths of the peak(s); $\sigma_{\rm cl}$ determines the width of the spectrum; $v_{\rm ICM}$ determines the location of the main peak and the trough; $\Delta v$ shifts the spectrum in the velocity dimension. In Figure~\ref{fig:vary}, we illustrate the impact of each parameter on the emergent model spectrum. Each parameter was varied individually while others were kept fixed to the best-fit parameter values. 

Specifically, spectra 1 and 2 (near the LBG C15) are clearly single-peaked, although spectrum 1 has a subdominant red bump. Interestingly, although spectrum 1 appears to be broader and has a larger $\sigma_{\rm Ly\alpha}$ on the moment map, it actually requires a smaller $\sigma_{\rm {\rm cl}}$ ($\sim$\,150\,km\,s$^{-1}$) than spectrum 2. The fit of spectrum 1 has a very large $v_{\rm cl}$ ($\sim$\,800\,km\,s$^{-1}$) and a negligible $v_{\rm ICM}$, whereas the fit of spectrum 2 has comparable $v_{\rm cl}$ and $v_{\rm ICM}$ (both $\sim$\,500\,km\,s$^{-1}$). This is due to the fact that multiphase outflows increase the asymmetry as well as the width of the spectra, and this degeneracy is not captured by the moment analysis. In particular, for spectrum 1, the optical depth at the \lya\ line center $\tau_{0,\rm ICM}$\footnote{Here $\tau_{0,\rm ICM}$ = $n_{\rm HI,\,{\rm ICM}}r_{\rm gal}\sigma_{\rm HI}(T_{\rm ICM}, v_{\rm ICM})$, where $\sigma_{\rm HI}(T_{\rm ICM}, v_{\rm ICM})$ is the \lya\ absorption cross-section at \HI\ temperature $T_{\rm ICM}$ and velocity $v_{\rm ICM}.$}$\sim 1$, so the \lya\ photons likely interact with the ICM prior to the clumps (as opposed to spectrum 2 where $\tau_{0,\rm ICM}\ll 1$). This implies that photons can then scatter approximately orthogonally off the clumps (see Appendix~\ref{sec:fccrit}), which yields an additional broadening of the spectrum that is larger than $\sigma_{\rm cl}$. The derived $\Delta v$ ($\sim$\,--200\,km\,s$^{-1}$) is also consistent with the C15 systemic redshift ($z$ = 3.0975) measured from \oiii. Also note that although spectrum 1 exhibits a blue dominant peak and a smaller red `hump' (which is commonly interpreted as a signature of inflows), our outflow model has successfully reproduced the observed line profile.

Spectrum 3 has a fairly narrow main peak, which yields the lowest $\sigma_{\rm {\rm cl}}$ ($\sim$\,100\,km\,s$^{-1}$) of all eleven sampled spectra. The fit has comparable $v_{\rm cl}$ and $v_{\rm ICM}$ (both $\sim$\,350\,km\,s$^{-1}$) and a small $\Delta v$ ($\sim$\,50\,km\,s$^{-1}$). It also has some dubious emission on the far blue side, which is not captured by the clumpy models. Increasing $\sigma_{\rm cl}$ in order to include both the red and blue peaks could potentially provide a better fit but would lack physical motivation. 

Spectra 4 and 5 (the tail of the tadpole) both possess two comparable peaks and a trough in the middle. The best-fit of the former is single-peaked, while the latter is double-peaked and captured the trough. Both fits have very large $\sigma_{\rm {\rm cl}}$ (>\,500\,km\,s$^{-1}$) to account for the line widths. The fit of spectrum 4 has comparably large $v_{\rm cl}$ and $v_{\rm ICM}$ (both $\sim$\,700\,km\,s$^{-1}$), whereas spectrum 5 has a $v_{\rm cl}$ of $\sim$\,500\,km\,s$^{-1}$ and a negligible $v_{\rm ICM}$.

Spectra 6 and 7 (the body of the tadpole) are both multi-peaked and dominated by a red peak. The best-fits both have $\sigma_{\rm {\rm cl}}\sim$\,450\,km\,s$^{-1}$. They have moderate $v_{\rm cl}$ ($\sim$\,200 and 400\,km\,s$^{-1}$, respectively) and small $v_{\rm ICM}<50$\,km\,s$^{-1}$ (dictated by the location of the absorption features).

Spectra 8 and 9 (the head of the tadpole, near ALMA-a) are both very broad ($\sigma_{\rm cl}$ $\sim$\,500\,km\,s$^{-1}$) and red-dominant double-peaked with a deep trough between two peaks. They both have high $F_{\rm V}$ ($\sim$\,0.5) and moderate $v_{\rm cl}$ ($\sim$\,250\,km\,s$^{-1}$) and $v_{\rm ICM}$ ($\sim$\,400\,km\,s$^{-1}$).

Spectra 10 and 11 (near the LBG C11) are also red-dominant double-peaked, although with slightly narrower line widths ($\sigma_{\rm {\rm cl}}$ $\sim$\,400\,km\,s$^{-1}$). Compared with spectra 8 and 9, they have lower $F_{\rm V}$ (<\,0.3), comparable $v_{\rm cl}$ ($\sim$\,250\,km\,s$^{-1}$) and lower $v_{\rm ICM}$ (<\,300\,km\,s$^{-1}$). The derived $\Delta v$ of spectrum 10 ($\sim$\,--150\,km\,s$^{-1}$) is also consistent with the C11 systemic redshift ($z$ = 3.0980\,$\pm$\,0.0001) measured from \oiii. Notably, the prominent double peak profiles in this region require a considerable outflow velocity for both the clumps and the ICM. This strongly suggests the presence of outflows, which is consistent with the indication of the large global velocity offsets between \lya, \oiii\ and Si ${\textsc {ii}}$.

It is noteworthy that although the observed $v_{\rm Ly\alpha}$ is fairly small in many positions (e.g., spectra 4 -- 9), large $\sigma_{\rm {\rm cl}}$, $v_{\rm cl}$ and and non-zero $v_{\rm ICM}$ are still preferred by the broad, asymmetric \lya\ profiles\footnote{This is also notable in cases where an \oiii\ measurement is available and the velocity offset between \lya\ and \oiii\ is close to zero (e.g., spectra 1 and 2).}. This concerns us that the first moment does not fully capture the gas kinematic information encoded in the \lya\ profiles. Second moments may be helpful in quantifying the line widths, whose possible physical interpretation is the random velocity dispersion of \HI\ clumps. Furthermore, the outflow velocity (parameterized as $v_{\rm cl}$ and $v_{\rm ICM}$ in the model) may be difficult to determine directly from the observed spectra (especially for complex \lya\ profiles), but might be retrieved using realistic radiative transfer modeling. 

In addition, we note that both the average and the standard deviation of all the derived $\Delta v$ are fairly small ($\langle\Delta v\rangle$ = --122\,km\,s$^{-1}$, $\sigma (\Delta v)$ = 87\,km\,s$^{-1}$), despite the large outflow velocities indicated by many of the \lya\ profiles. This corresponds to an average systemic redshift of LAB1, $\langle z_{\rm sys}\rangle$ = 3.0983 $\pm$ 0.0004.

We caution that the effect of \lya\ absorption from the intergalactic medium (IGM) is not modeled in this work. It is expected that at $z\sim 3$ this effect is in general non-negligible on the blue side of the spectrum \citep{2007MNRAS.377.1175D,Laursen11}. However, we do not expect the effect of the IGM to be significant here, as it would cause sharp absorption troughs and yield multiple peaks, which should be clearly visible given the widths of the observed spectra (see \citealp{2020arXiv200610041B} for a discussion of this effect) instead of simply attenuating the spectrum smoothly. 

To recap, the main results of our analysis are:

\begin{enumerate}
    \item The observed \lya\ spectra require relatively few clumps per line-of-sight ($f_{\rm cl} \lesssim f_{\rm cl, crit}$) as they have significant fluxes at the line center. Therefore, they are very different from the spectra of most \lya\ emitting {\it galaxies} at essentially all redshifts (e.g., \citealt{Steidel10, Erb14, Trainor15, Yang16, Yang17a, Gronke17, Orlitova18}), which can usually be reproduced by a uniform medium (e.g., the `shell model')\footnote{The success of shell model fitting may no longer be achieved when the model parameters are further constrained by additional observations (e.g., optical emission lines or UV absorption lines, see \citealt{Orlitova18}).} or by a multiphase medium with a large number \HI\ clumps ($f_{\rm cl} \gg f_{\rm cl, crit}$). 
    
\item The velocity dispersion of the scattering clumps yields a broadening of the spectra from the intrinsic line width $\sigma_{\rm i}\sim 13$\,km\,s$^{-1}$ to $\gg 100$\,km\,s$^{-1}$ as observed. This is possible when $f_{\rm cl} \sim \alpha f_{\rm cl, crit}$ with $\alpha \sim$ a few\footnote{Note that the scattering off the surface of the clumps broadens the spectrum as long as $\sigma_{\rm cl}>\sigma_{\rm i}$ (and $f_{\rm cl} \sim f_{\rm cl, crit}$). Hence, the emergent spectra are insensitive to the exact value of $\sigma_{\rm i}$.}. Such a process may be crucial in galaxies where the observed \lya\ line is always broader, usually by at least a factor of two, than the corresponding non-resonant lines such as \oiii, \ha\ or \hb\ \citep[e.g.,][]{Orlitova18}.

\item While the widths of the spectra are set primarily by the velocity dispersion of the clumps, i.e., $\sigma_{\rm Ly\alpha}\sim \sigma_{\rm cl}$, we found that the clump bulk outflow can also cause additional broadening, as seen in spectrum 1. In this case, one might naively assume that the photons do not interact with the clumps due to their large velocity offsets ($v_{\rm cl}\gg \sigma_{\rm cl}$). However, if $\tau_{0, \mathrm{ICM}}\gtrsim$ 1, the photons may first interact with the ICM, which significantly reduces the parallel component of $v_{\rm cl}$ ($v_{{\rm cl}, \parallel}$) appearing to the photons, and hence greatly increases the optical depth. This result suggests that we may have interpreted our model too naively (e.g., using single-scattering approximation), especially considering that the kinematics of our model are clearly simplistic and not strictly hydrodynamically stable (we usually expect $v_{\rm cl}\sim v_{\rm ICM}$, i.e., the clouds are entrained by the local flow, see e.g., \citealt{1994ApJ...420..213K,2020MNRAS.492.1841L})\footnote{We do see $v_{\rm cl}\sim v_{\rm ICM}$ in our fits of spectra 3, 4, 5, 10 and 11.}. Moreover, although we found that significant outflow velocities ($\gtrsim$\,100\,km\,s$^{-1}$) are required to reproduce the observed spectra, the exact values may still be subject to considerable uncertainties, due to the internal degeneracies and the presumably more complicated kinematics in reality (e.g., \citealt{Steidel10} show that gas outflows even within the same galaxy have a range of velocities that goes from 0 to 800\,km\,s$^{-1}$ with varying effective optical depths).

\item In our best-fit spectra, the \HI\ in the ICM is responsible for the absorption feature close to the line center (cf. spectrum 5 or 6). However, several tentative absorption features can be present in a single spectrum (e.g., spectra 5, 6, 7 and 9), and they are not captured simultaneously by our model. These multiple features might be caused by the \HI\ in the outer CGM / IGM, where the probability of back-scattering into the line-of-sight is negligible.

\end{enumerate}

The derived values of our fitted parameters also fit into a broader picture in at least two ways:

\begin{enumerate}

\item The $\sigma_{\rm cl}$ values correspond to reasonable dark matter halo masses. The dynamical mass of the LAB1 halo can be estimated from the velocity dispersion and physical size (assuming spherical symmetry):

\begin{equation}
M_{\rm dyn} = \frac{3\sigma_{\rm cl} ^{2}R}{G} = 6.9\times10^{9}
\left(\frac{\sigma_{\rm cl}}{100\ \mathrm{km\ s}^{-1}}\right)^2\left(\frac{R}{\mathrm{kpc}}\right)
M_{\odot}. 
\end{equation}

Taking $R$ $\sim$\,100 kpc, the highest $\sigma_{\rm cl}$ ($\sim$\,700\,km\,s$^{-1}$) corresponds to $M_{\rm dyn}$ $\sim$ $10^{13.5}$ $M_{\odot}$. This result is consistent with the predicted halo masses from the Millennium simulations at $z$ = 3.06. As calculated by \citet{Kubo16}, the halo masses range from $10^{12.2}$ to $10^{14.0}$ $M_{\odot}$ (with median $10^{13.2}$ $M_{\odot}$) for the \citet{Lucia07} model, and from $10^{12.4}$ to $10^{14.1}$ $M_{\odot}$ (with median $10^{13.2}$ $M_{\odot}$) for the \citet{Guo11} model.

\item The $v_{\rm cl}$ values correspond to reasonable survival times of the clumps. Here we consider two different criteria proposed by \citet{2020MNRAS.492.1841L} and \citet{Gronke2018}, respectively. Following \citet{2020MNRAS.492.1841L}, assuming the ionized medium has \HII\ number density $n_{\rm hot}$ = 0.01 $\rm cm^{-3}$, an outflow velocity $v_{\rm cl}$ $\sim$ 500 km\,$\rm s^{-1}$ corresponds to a cloud lifetime $t_{\rm life}$ $\sim$\,100 Myr. The cooling time of the hot medium, $t_{\rm cool, h}$ $\sim$\,30 Myr $\leq$ $t_{\rm life}$. Whereas following \citet{Gronke2018}, the cooling time of the mixing layer $t_{\rm cool, mix}$ $\sim$\,3 Myr, and the cloud-crushing time $t_{\rm cc}$ $\sim$\,1 Myr $\simeq$ $t_{\rm cool, mix}$, which implies possible survival of the cold gas. So either criterion indicates that the clumps can survive for a fairly long time, and may even grow in mass as they accrete the cooling hot material from the ambient medium.

\end{enumerate}

Recently, \citet{Herenz20} have reported a significant detection of {\rm He\,{\textsc {ii}}}\,$\lambda$1640 emission in three regions of LAB1 (which are close to C15, the tail of the tadpole structure, and C11, respectively) as well as a non-detection of {\rm C\,{\textsc {iv}}}\,$\lambda\lambda$1548, 1550 doublet. They have carried out a detailed analysis and concluded that their observed {\rm He\,{\textsc {ii}}}/\lya\ and {\rm C\,{\textsc {iv}}}/\lya\ ratios are consistent with cooling radiation, feedback driven shocks, and/or photo-ionisation from an embedded AGN. We examined our MOSFIRE spectra around these regions but did not find additional rest-UV collisionally excited emission lines near \lya\ that are significant, although we do see significant outflow velocities in these three regions (cf. spectra 1, 2, 5, 10, and 11). More observations are needed to distinguish these different powering mechanisms.

In summary, the multiphase clumpy model is versatile enough to reproduce the diverse \lya\ morphologies observed. The fitting results are still, not surprisingly, model dependent -- different assumptions on the geometry and moving pattern of the \HI\ gas may yield different results. Furthermore, our modeling with parameters of scattering medium varying independently at different locations can be handled in a more self-consistent manner, as in reality the \lya\ photons are likely to be scattered by a common, spatially varying medium. Nonetheless, our analysis is a first attempt to model the spatially-resolved \lya\ profiles in LAB1 with more physically realistic clumpy models. It provides us with insights on the gas kinematics and will serve as the foundation of more advanced radiative transfer modeling in the future. One promising future direction is to use more elaborate clump velocity profiles (e.g., consistent with absorption line observations) which can alter $f_{\rm cl,crit}$ (cf. Appendix~\ref{sec:fccrit}). We will explore such new physical regimes in our future work.

\section{Conclusions}\label{sec:conclusion}
We have carried out deep spectroscopic observations of SSA22-LAB1 at $z$ = 3.1 using KCWI and MOSFIRE. The main conclusions of our analysis are:
\begin{enumerate}
\item By applying matched filtering to the KCWI datacube, we have created a narrow-band \lya\ image of LAB1. The most prominent feature is a tadpole-shaped structure, whose `head' overlaps with one of the ALMA sources yet whose `tail' does not associate with any identified sources; 

\item By comparing the spatial distributions and intensities of \lya\ and \hb, we find that recombination of photo-ionized \HI\ gas followed by resonant scattering is sufficient to explain all the observed \lya/\hb\ ratios;

\item Using both moment map analysis and MCRT modeling, we have managed to extract physical information from the spatially-resolved \lya\ profiles. We find that moment maps can be used as a crude indicator of the \HI\ gas kinematics, but realistic MCRT modeling needs to be invoked to extract detailed kinematic information and make physical interpretations. By fitting a set of multiphase, `clumpy' models to the observed \lya\ profiles, we are able to reasonably constrain many physical parameters, namely the \HI\ number density in the ICM, the cloud volume filling factor, the random velocity and outflow velocity of the clumps, the \HI\ outflow velocity of the ICM and the local systemic redshift. Our model has successfully reproduced the diverse \lya\ morphologies at different locations, and the main results are: (1) The observed \lya\ spectra require relatively few clumps per line-of-sight ($f_{\rm cl} \lesssim f_{\rm cl, crit}$) as they have significant fluxes at the line center; (2) The velocity dispersion of the scattering clumps yields a significant broadening of the spectra as observed; (3) The clump bulk outflow can also cause additional broadening if $\tau_{0, \mathrm{ICM}}\gtrsim$ 1. In that case, the photons may first interact with the ICM, which significantly reduces the parallel component of clump outflow velocity appearing to the photons, and hence greatly increases the optical depth of the clumps; (4) The \HI\ in the ICM is responsible for the absorption feature close to the \lya\ line center.
\end{enumerate}

We caution that there are still a number of caveats to this study. For example, our MCRT modeling is inherently model dependent, in particular on the specific assumptions about the kinematics of the cold clumps. A combination of results from hydrodynamical simulations and additional observations (e.g., absorption line studies) may help constrain the actual gas kinematics better. We intend to explore these possibilities in our future work. 

\section*{Acknowledgements}
We thank the referee for insightful comments that have significantly improved the paper. We also thank Phil Hopkins for providing us with adequate computational resources. ZL acknowledges Michael Zhang, for his professional guidance and kind company during the excruciating debugging process. The data presented herein were obtained at the W. M. Keck Observatory, which is operated as a scientific partnership among the California Institute of Technology, the University of California and the National Aeronautics and Space Administration. The Observatory was made possible by the generous financial support of the W. M. Keck Foundation. We are also grateful to the dedicated staff of the W.M. Keck Observatory who keep the instruments and telescopes running effectively. MG was supported by NASA through the NASA Hubble Fellowship grant HST-HF2-51409 awarded by the Space Telescope Science Institute, which is operated by the Association of Universities for Research in Astronomy, Inc., for NASA, under contract NAS5-26555. Numerical calculations were run on the Caltech compute cluster ``Wheeler,'' allocations from XSEDE TG-AST130039 and PRAC NSF.1713353 supported by the NSF, and NASA HEC SMD-16-7592. This research made use of Montage. It is funded by the National Science Foundation under Grant Number ACI-1440620, and was previously funded by the National Aeronautics and Space Administration's Earth Science Technology Office, Computation Technologies Project, under Cooperative Agreement Number NCC5-626 between NASA and the California Institute of Technology.

\section*{Data availability}
The data underlying this article will be shared on reasonable request to the corresponding author. 

\bibliographystyle{mnras}
\bibliography{LAB1}

\begin{appendix}
\section{Gaussian fits to the [OIII] profiles of LBGs C11 and C15}\label{sec:C11_C15}
In order to determine the systemic redshifts of two LBGs, C11 and C15, we fitted the \oiii\ profiles (5008.24\,\AA) with single Gaussians. The \oiii\ emission is spatially integrated over the ranges indicated by black solid arrows in Figure \ref{fig:OIII_Slits}, which include all the significant \oiii\ emission of C11 and C15.

We used the \texttt{PySpecKit} package \citep{Ginsburg11} to fit the \oiii\ profiles with a single Gaussian model:

\begin{equation}
    \label{eq:gaussian}
    F_{\rm \lambda} = F_{\rm 0}\,e^{-\frac{(\lambda - \lambda_{0})^2}{2\sigma^2}} 
\end{equation}

\begin{figure*}
\centering
\includegraphics[width=0.80\textwidth]{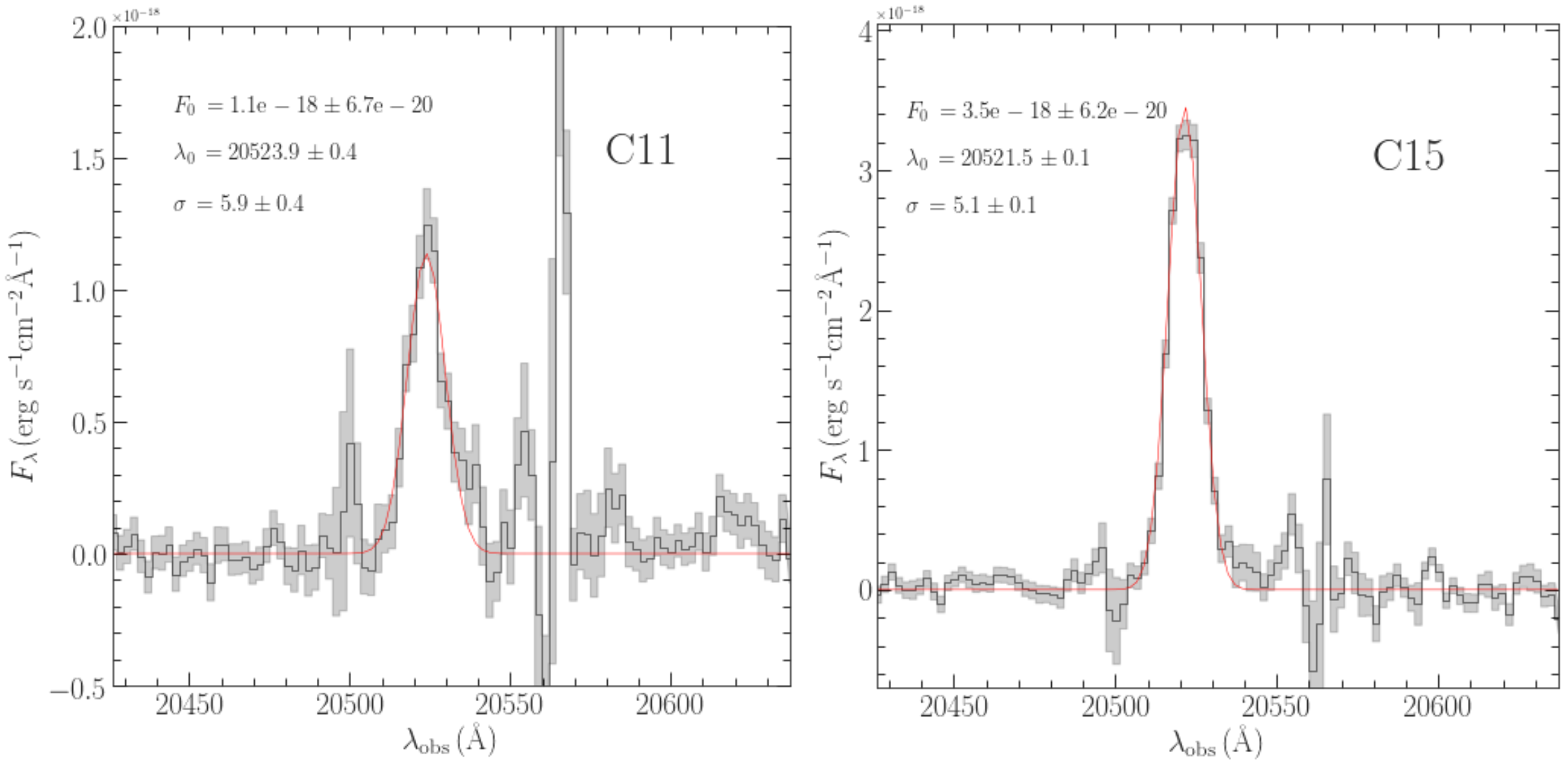}
    \caption{Single Gaussian fits to the observed \oiii\ profiles of two LBGs, C11 and C15. The \oiii\ emission is spatially integrated over the ranges indicated by black solid arrows in Figure \ref{fig:OIII_Slits}, which include all the significant \oiii\ emission of C11 and C15. The observed fluxes (shown in black) with 1-$\sigma$ uncertainties (shaded in grey), the best-fit model (shown in red) and best-fit parameters (with 1-$\sigma$ uncertainties) are shown in each panel. Note that skylines are present near the \oiii\ lines, as indicated by large flux uncertainties.
    \label{fig:c11c15}}
\end{figure*}

The fitting results and derived values with 1-$\sigma$ uncertainties of the free parameters ($F_0, \lambda_0$ and $\sigma$) are shown in Figure \ref{fig:c11c15}. The systemic redshifts of C11 and C15 are therefore determined to be \emph{z}(C11) = 3.0980\,$\pm$\,0.0001, \emph{z}(C15) = 3.0975\,$\pm$\,0.0001 (accounting for the typical redshift precision achieved by MOSFIRE measurements reported in \citealt{Steidel14}). 

\section{Derivation of the critical clump covering fraction}\label{sec:fccrit}
Here we analytically derive the critical clump covering fraction, $f_{\rm cl, crit}$, for a multiphase medium whose kinematics have been defined earlier in this work, i.e., \HI\ clumps with a constant outflow velocity $v_{\rm cl}$ and a velocity dispersion $\sigma_{\rm cl}$, and an inter-clump medium (ICM) with a constant outflow velocity $v_{\rm ICM}$. The optical depth of the \HI\ in the ICM at the \lya\ intrinsic frequency is approximated as $\tau_{0,{\rm ICM}}$, i.e., the optical depth at the line center\footnote{Technically speaking, the optical depth of the \HI\ in the ICM is expressed as $\int\mathrm{d}v\,f_{\rm i}(v)\tau_{\rm ICM}(v)$ where $f_{\rm i}(v)$ is the normalized intrinsic spectrum as a function of velocity. As our intrinsic spectrum is very narrow, this approximation holds.}.

The large widths of the observed spectra imply that photons escape in a single long flight -- after they have been scattered off the surface of a fast moving clump \citep{Gronke17b}. It follows that $f_{\rm cl, crit}$ is given by the condition that on average one clump interacts with a photon at its intrinsic frequency.

The width in velocity space of each clump, $\tilde v$, is determined by:
\begin{equation}
    \label{eq:vtilde}
    \tau_{\rm cl}(\pm \tilde v / 2)= \frac{4}{3} N_{\rm HI, cl} \sigma_{\rm HI}(\pm \tilde v / 2) = 1
\end{equation}
where $\tau_{\rm cl}$ is the optical depth of the clump, $N_{\rm HI, cl}$ is the \HI\ column density of the clump, and $\sigma_{\rm HI}$ is the \lya\ cross section of the clump. The factor $4/3$ is simply due to the spherical geometry of the clump. 
Using the core approximation of the \lya\ cross section ($\sigma_{\rm cl}(v)\propto \exp(-v^2/v_{\rm th}^2)$), the solution to Eq. (\ref{eq:vtilde}) can be explicitly written as
$\tilde v = 2 v_{\rm th} \sqrt{\ln\tau_{\rm 0, cl}}$, where $\tau_{\rm 0, cl}\equiv \tau_{\rm cl}(v=0)$, and $v_{\rm th}$ is the thermal velocity dispersion of \HI\ within the clumps. For the \HI\ column density and temperature of the clumps used in this work, $\tilde v \simeq$ 78\,km\,s$^{-1}$.

Under the assumption that all photons are injected at the \lya\ line center (which is a reasonable approximation for our setup since the width of the intrinsic spectrum $\sigma_{\rm i} \ll \sigma_{\rm cl}$), the average number of clumps per line-of-sight that intersect with $v=0$ in velocity space is:
\begin{equation}
    \label{eq:fctilde}
    \tilde f_{\rm cl} = f_{\rm cl}\int\limits_{-\tilde v}^{\tilde v}\mathcal{N}(v, \mu = v_{{\rm cl}, \parallel}, \sigma = \sigma_{\rm cl})\,\mathrm{d}v
\end{equation}
where $\mathcal{N}$ denotes the normal distribution assumed for the velocity distribution of the clumps\footnote{This choice is purely for practical purpose -- in principle it can be replaced by any physically reasonable velocity distribution.}. Here $v_{{\rm cl}, \parallel}$ is a component of $v_{\rm cl}$ that is parallel to the trajectories of photons. The reason for considering $v_{{\rm cl}, \parallel}$ rather than $v_{\rm cl}$ is explained below.

Given the considerations above, demanding $\tilde f_{\rm cl} = 1$ yields the critical number of clumps per line-of-sight (i.e., the clump covering fraction):
\begin{equation}
    \label{eq:fccrit}
    f_{\rm cl, crit} =  \frac{2}{{\rm erf} \left( \frac{-v_{{\rm cl},\parallel} + 2v_{\rm th}\sqrt{\ln\tau_{0,\rm cl}} }{\sqrt{2}\sigma_{\rm cl}}\right) + {\rm erf} \left( \frac{v_{{\rm cl},\parallel} + 2v_{\rm th}\sqrt{\ln\tau_{0,\rm cl}} }{\sqrt{2}\sigma_{\rm cl}}\right)}\;.
\end{equation}
where $\mathrm{erf}(x)$ is the Gauss error function. 

Note that this equation is a generalization of equation (12) in \citet{Gronke17b}, where the radial velocity distribution of the clumps is approximated as a tophat profile. It converges to that equation when $v_{{\rm cl},\parallel}\ll \sigma_{\rm cl}$. For $v_{{\rm cl},\parallel}\gg \sigma_{\rm cl}$, this equation yields very large values of $f_{\rm cl,crit}$.

Here we discuss $v_{{\rm cl},\parallel}$ for two different cases: (1) If $\tau_{0,{\rm ICM}} \ll 1$, initially the photons do not interact with the ICM. Therefore, $v_{{\rm cl},\parallel}\approx v_{\rm cl}$; (2) If $\tau_{0, {\rm ICM}} \sim 1$ (up to a few), the photons can interact with the ICM prior to the clumps, and thus are likely to scatter orthogonally to the clump bulk outflow. Therefore, $v_{{\rm cl},\parallel}\approx 0$ (with details depending on the exact value of $\tau_{0,{\rm ICM}}$ and the clump distribution). We do not consider $\tau_{0, {\rm ICM}} \gg 1$ cases, where the multiphase, clumpy medium converges to a homogeneous medium that would fail to reproduce the observed spectra presented in this work (cf. discussion in \S\ref{sec:RT}).

\section{Justification of using spatially-integrated models to fit the spatially-resolved \lya\ profiles}\label{sec:justify}

In this work, we have fitted spatially-resolved \lya\ profiles with spatially-integrated models, which are derived from \lya\ photons from the entire scattering region. Here we emphasize that qualitatively, these spatially-integrated models are very similar to the binned models, which are derived from \lya\ photons within a certain impact parameter range. 

\begin{figure*}
\centering
\includegraphics[width=0.84\textwidth]{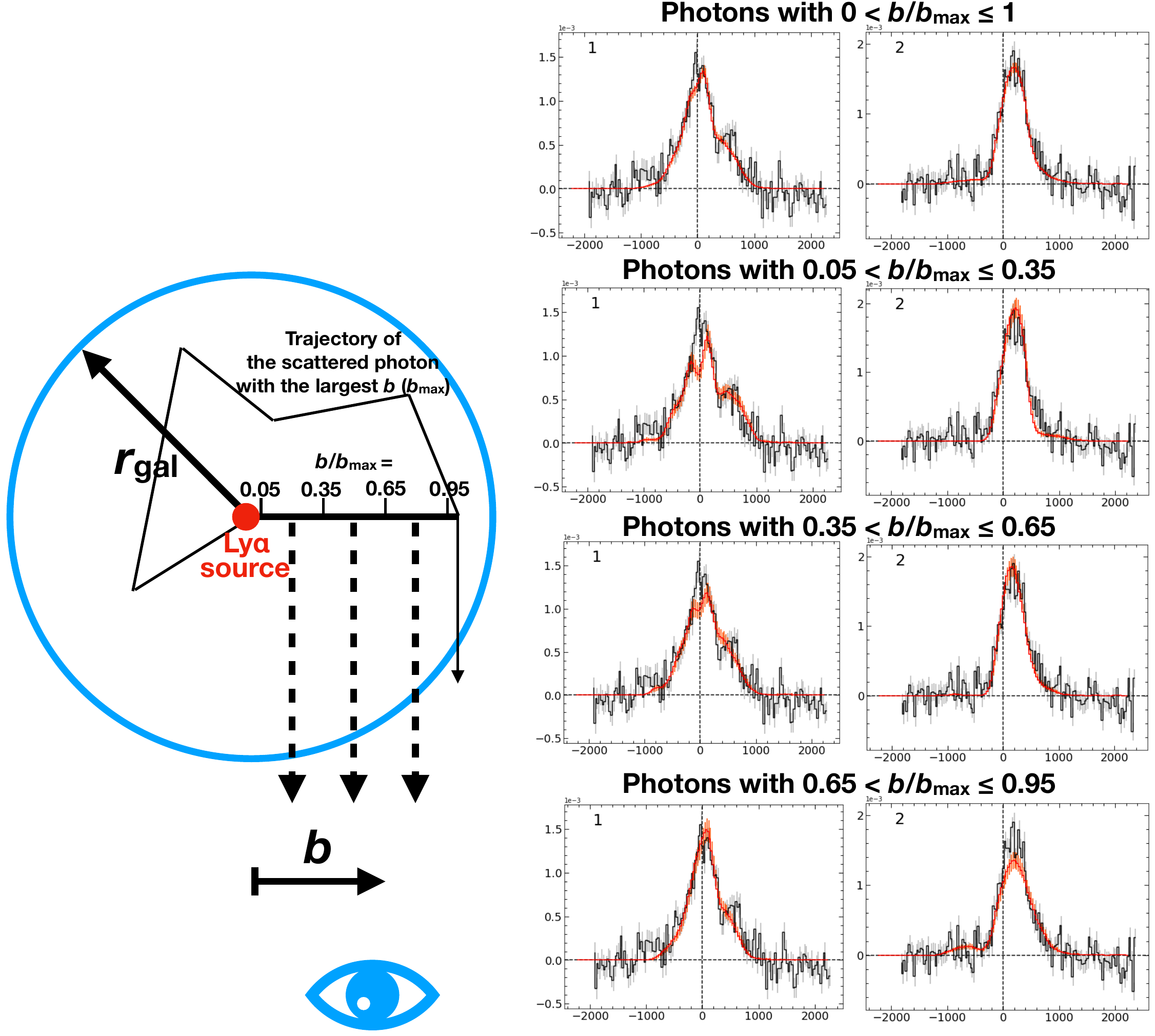}
    \caption{Justification of using spatially-integrated models to fit the spatially-resolved \lya\ profiles. \emph{Left}: Configuration of the multiphase, clumpy model and the way we construct our photon bins for the observer. \emph{Right:} Comparison of integrated models and binned models for spectra 1 and 2. Assuming the largest impact parameter of all the photons is $b_{\rm max}$, we made three photon bins within the 2$\sigma$ range ($0.05 < b/b_{\rm max} \leq 0.35$, $0.35 < b/b_{\rm max} \leq 0.65$ and $0.65 < b/b_{\rm max} \leq 0.95$) and constructed three binned model spectra with these photon bins respectively. It can be seen the binned models are qualitatively very similar to the integrated model.
    \label{fig:impact}}
\end{figure*}

We illustrate this in Figure \ref{fig:impact}. With spectra 1 and 2 as two examples, we binned all the scattered photons according to their impact parameters, $b$ (the projected distance to the simulation center perpendicular to the line-of-sight). Assuming the largest impact parameter of all the photons is $b_{\rm max}$, we made three photon bins within the 2$\sigma$ range ($0.05 < b/b_{\rm max} \leq 0.35$, $0.35 < b/b_{\rm max} \leq 0.65$ and $0.65 < b/b_{\rm max} \leq 0.95$) and constructed three binned model spectra with these photon bins respectively. It can be seen the binned models are qualitatively very similar to the integrated model. Therefore, it is approximately correct to model the spatially-resolved profiles at positions away from the \lya\ emitting sources with spatially-integrated models.

The aim of our modeling in this work is to roughly extract the velocities and densities of \HI\ that the \lya\ photons `experience' in-situ. In our following work, we plan to model the spatially-resolved profiles in a more self-consistent way (e.g., modeling the spatially-resolved \lya\ profiles with a common scattering medium). 

\section{Posterior Probability Distributions Derived From Nested Sampling}\label{sec:posterior}

\begin{figure*}
\centering
\includegraphics[width=\textwidth]{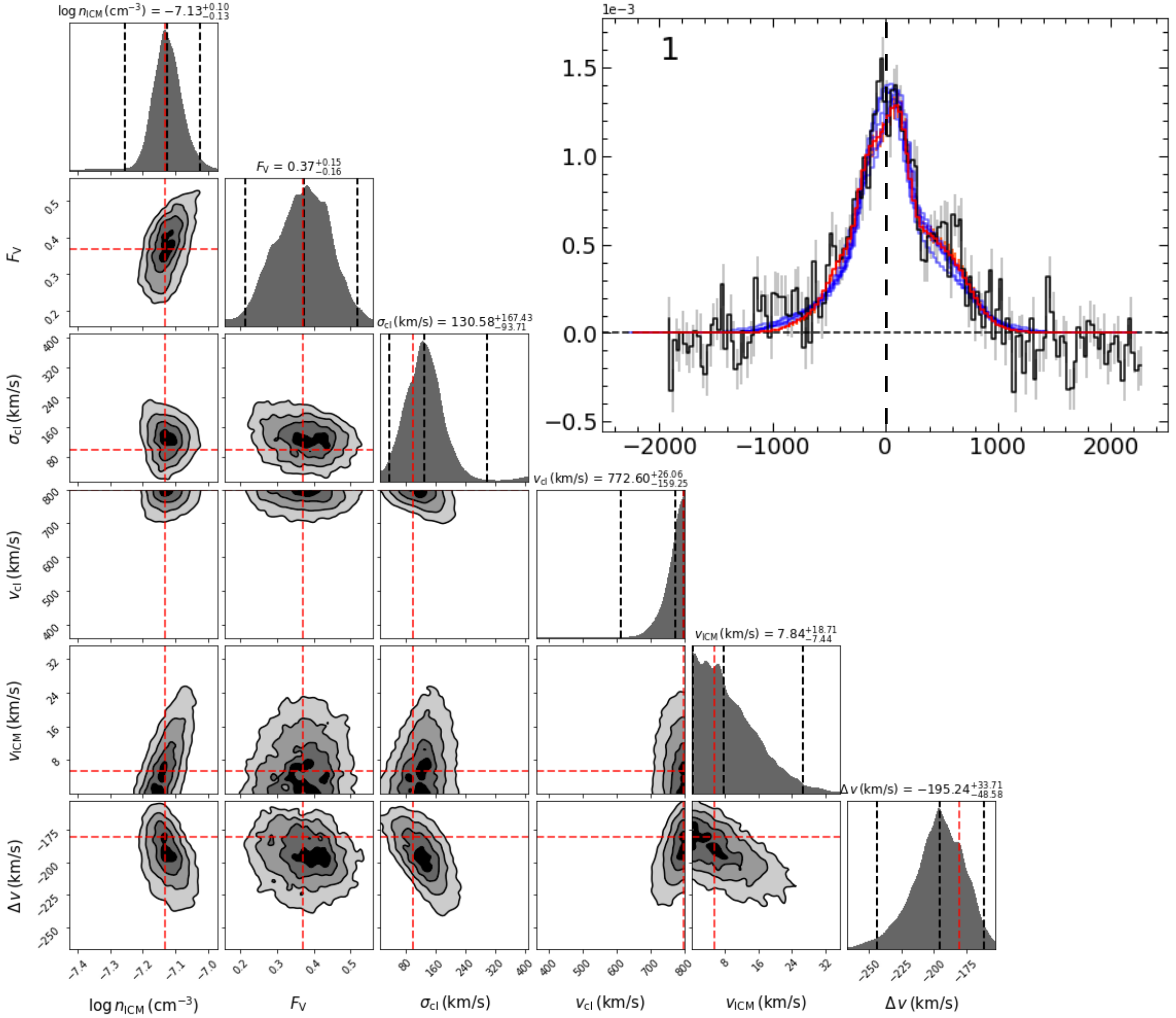}
\end{figure*}

Here we present the joint and marginal posterior probability distributions of the multiphase clumpy model parameters for all eleven representative \lya\ spectra derived from nested sampling. For each spectrum, we also show the best-fit, the observed \lya\ profile and five model spectrum samples from nested sampling.

\begin{figure*}
\centering
\includegraphics[width=\textwidth]{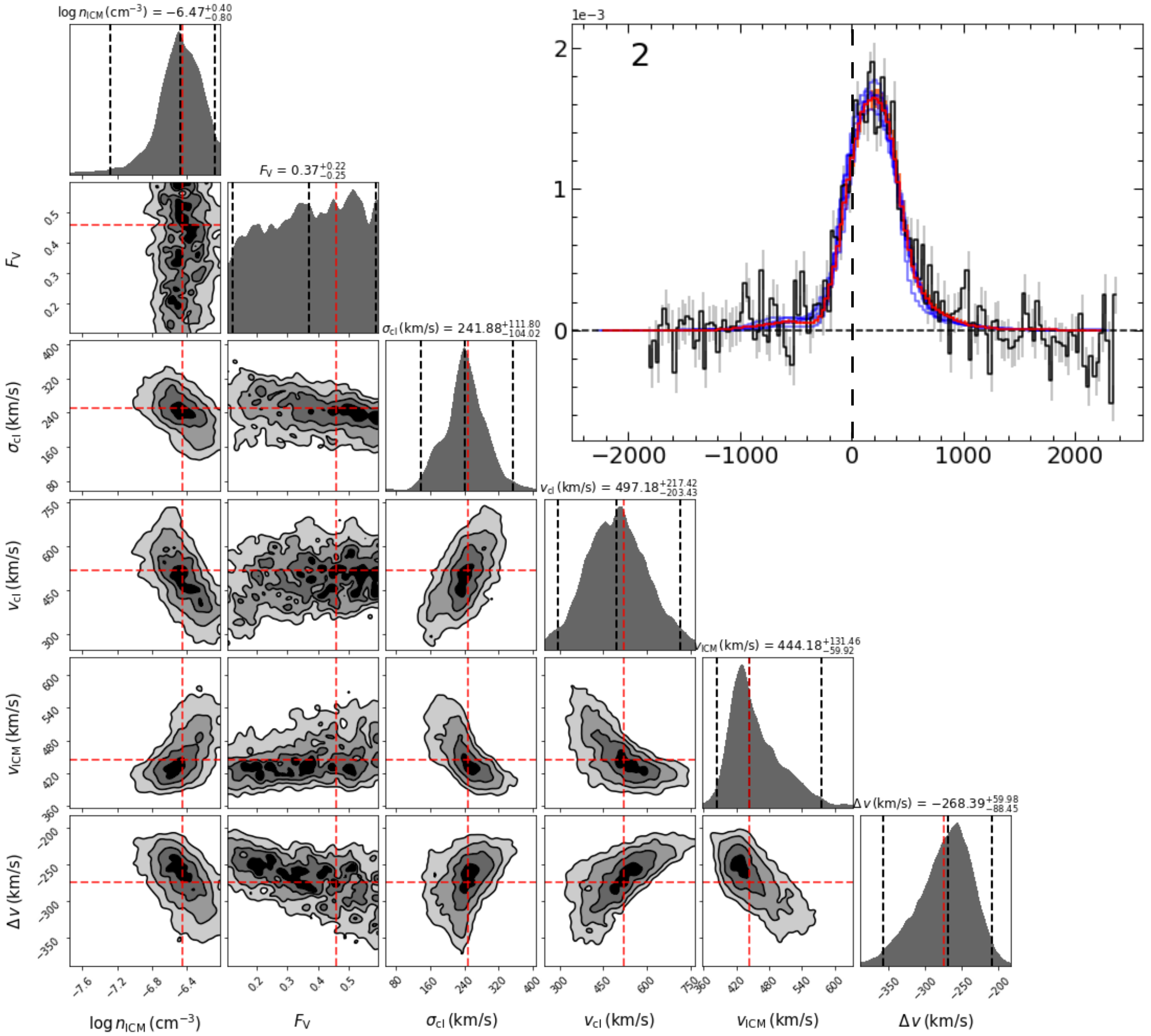}
\end{figure*}

\begin{figure*}
\centering
\includegraphics[width=\textwidth]{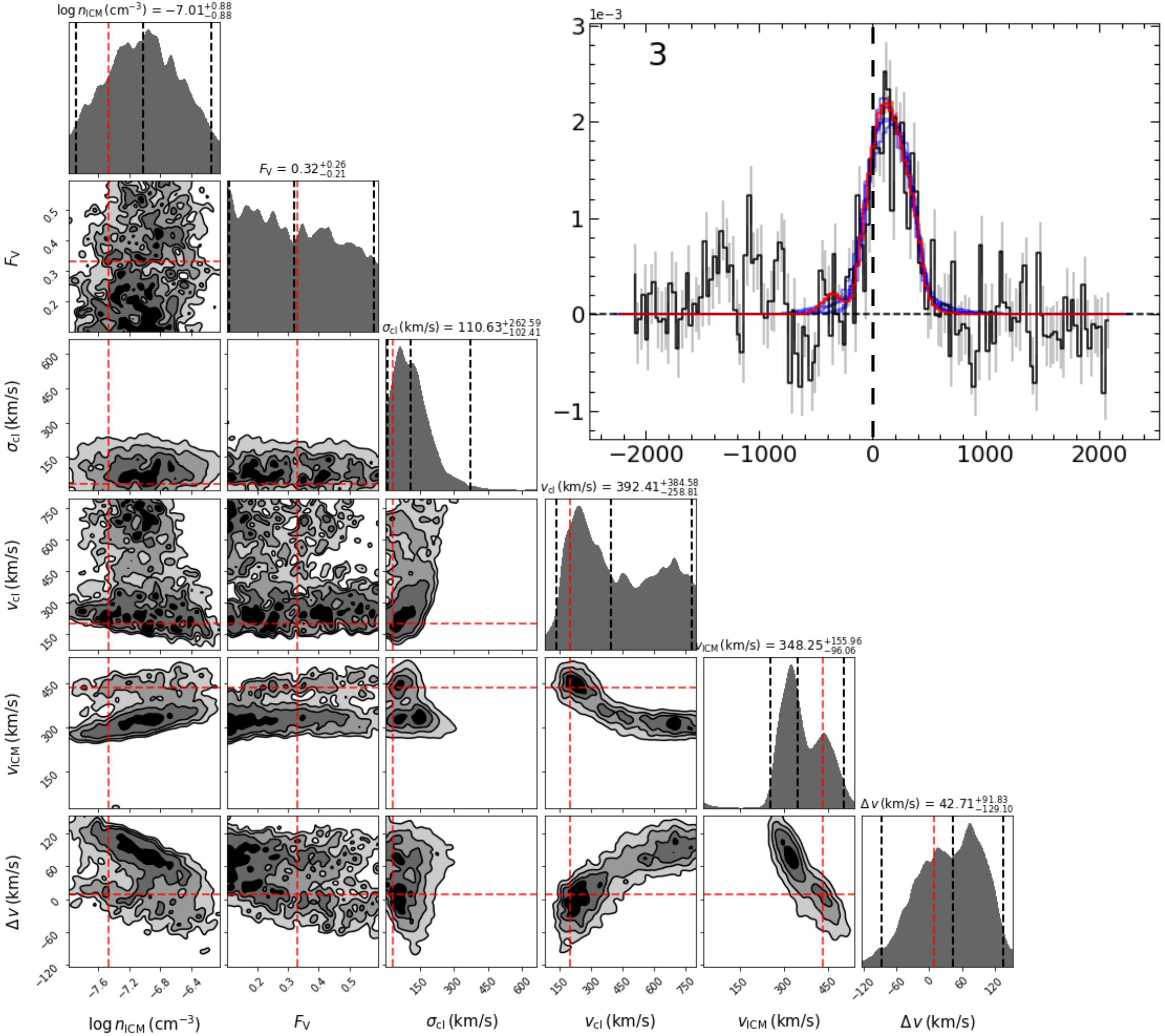}
\end{figure*}

\begin{figure*}
\centering
\includegraphics[width=\textwidth]{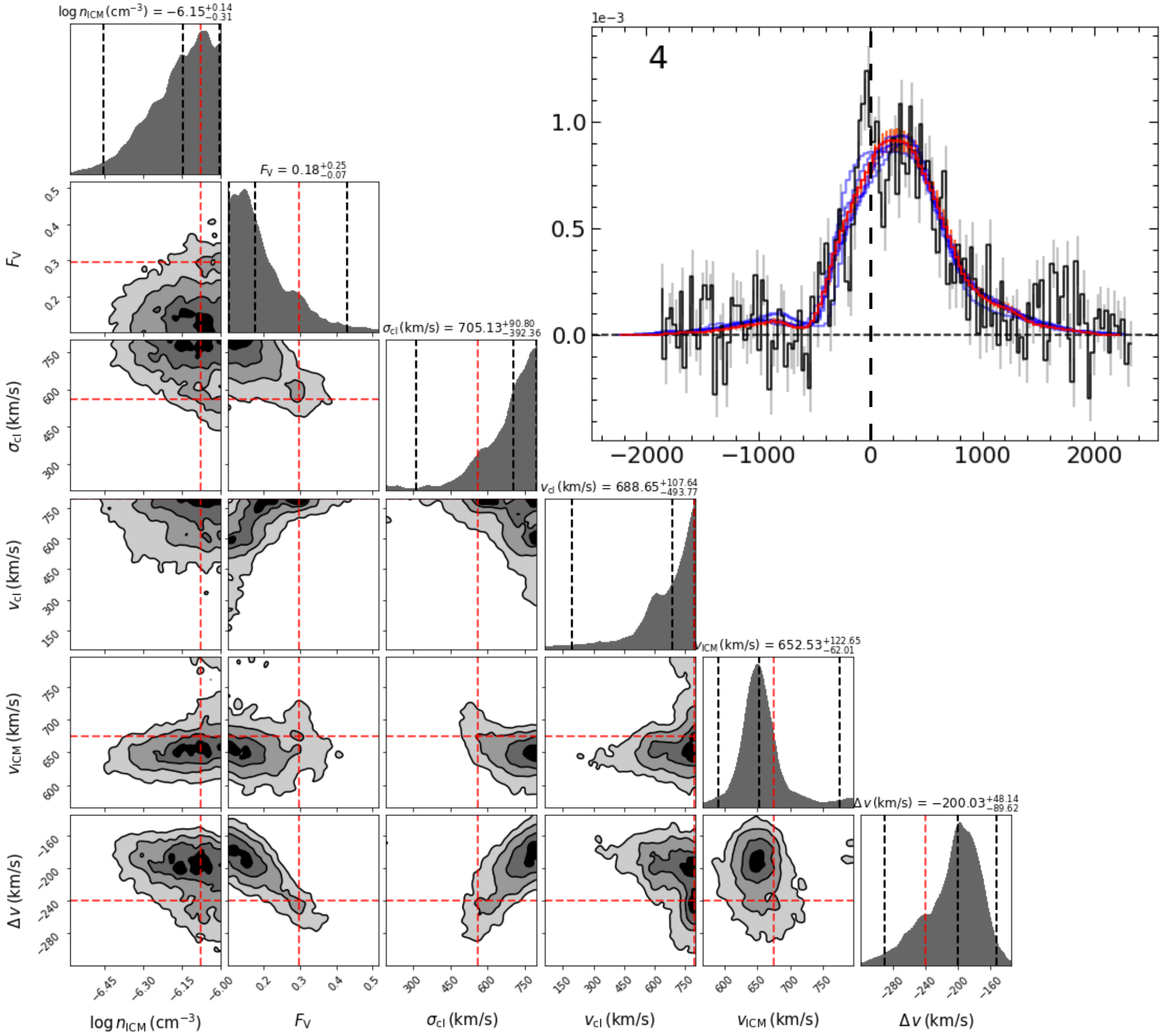}
\end{figure*}

\begin{figure*}
\centering
\includegraphics[width=\textwidth]{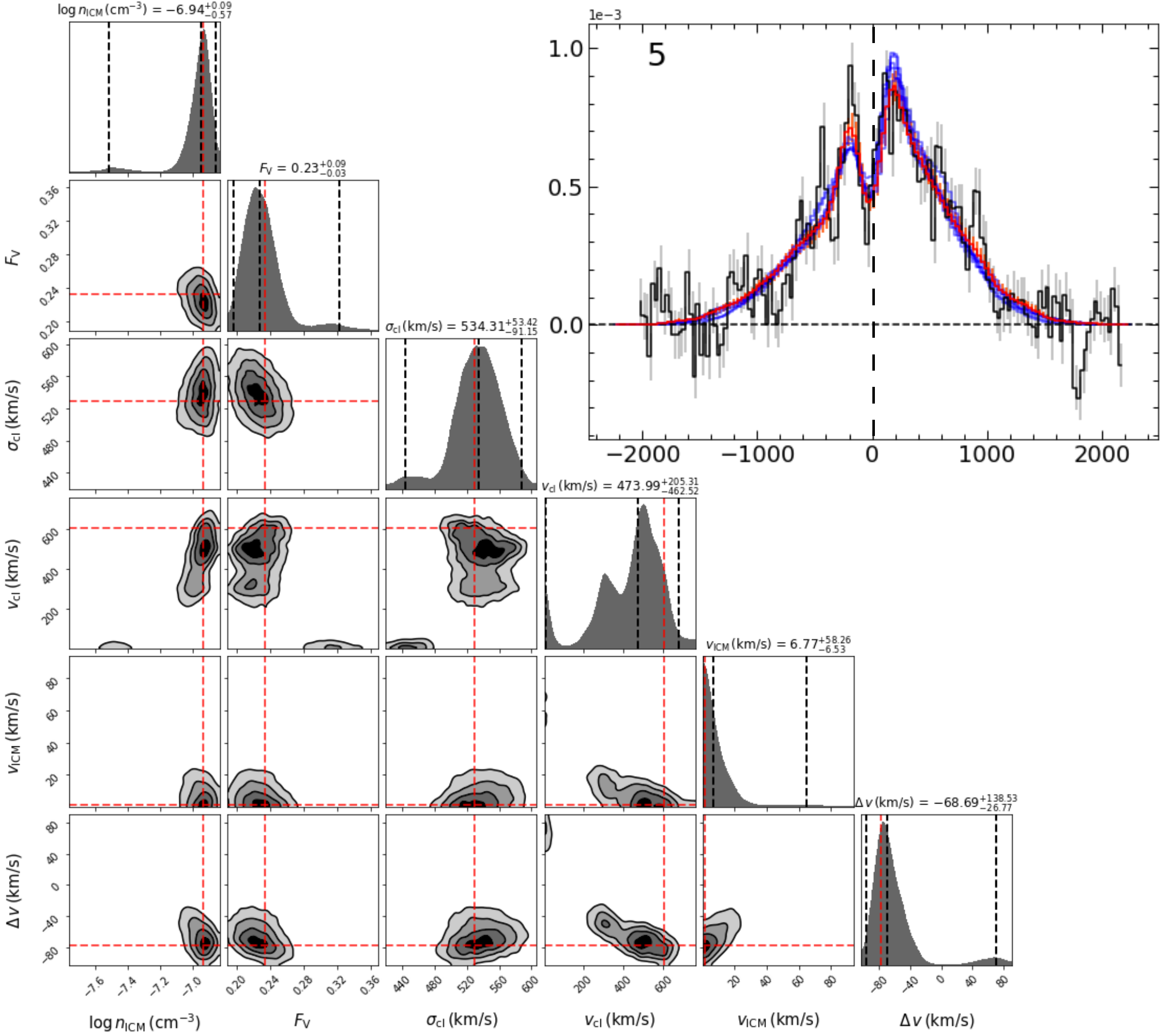}
\end{figure*}

\begin{figure*}
\centering
\includegraphics[width=\textwidth]{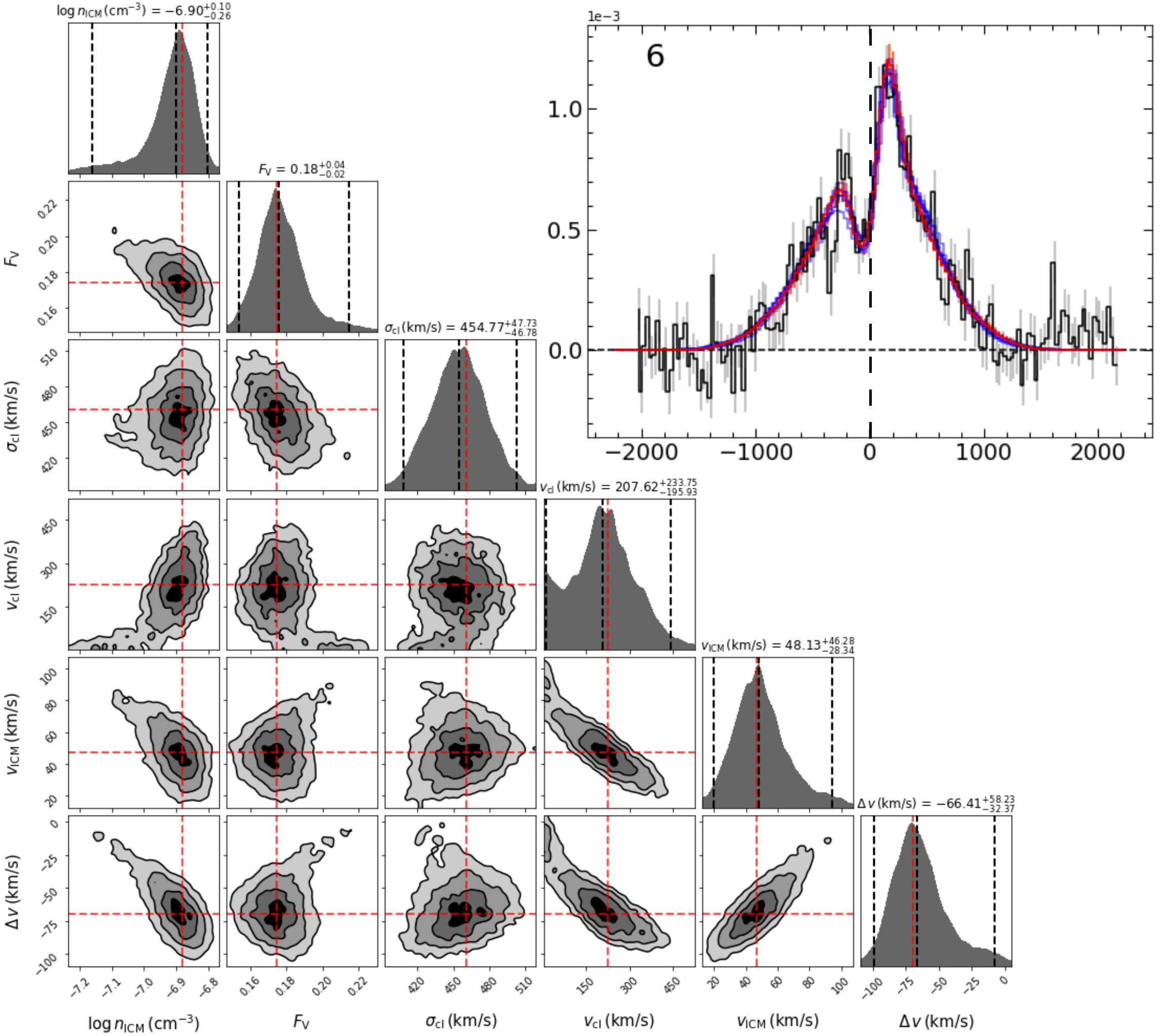}
\end{figure*}

\begin{figure*}
\centering
\includegraphics[width=\textwidth]{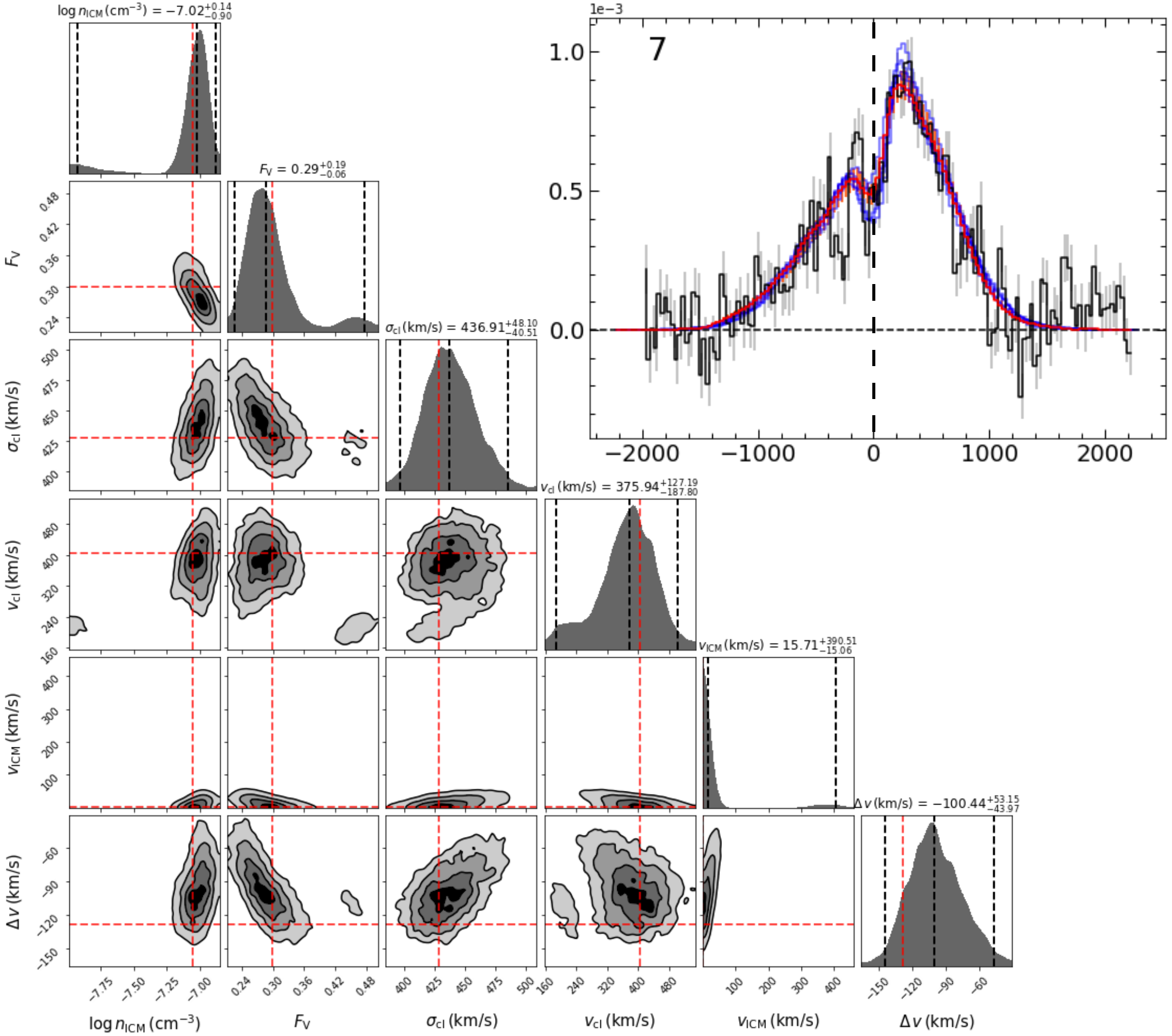}
\end{figure*}

\begin{figure*}
\centering
\includegraphics[width=\textwidth]{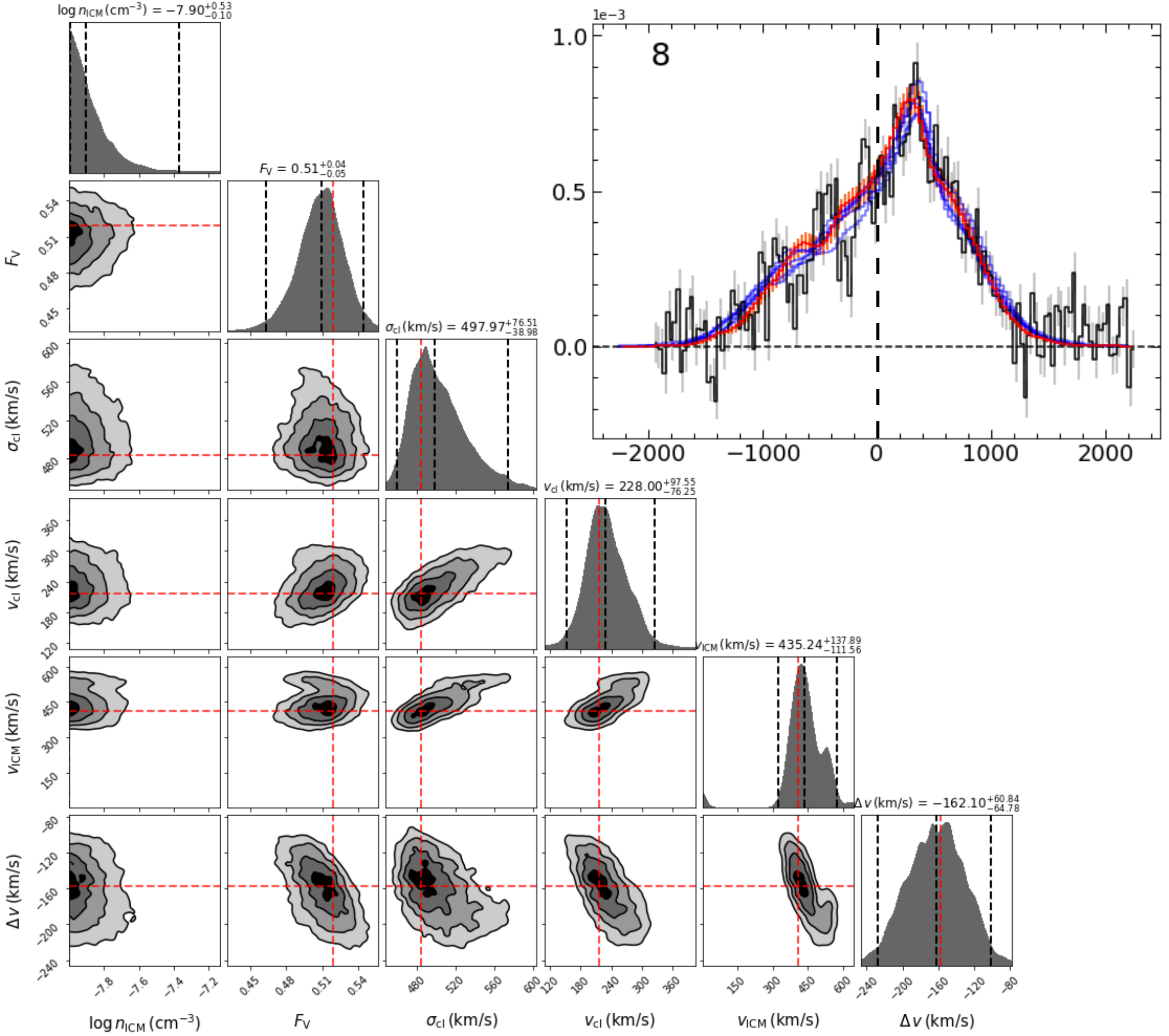}
\end{figure*}

\begin{figure*}
\centering
\includegraphics[width=\textwidth]{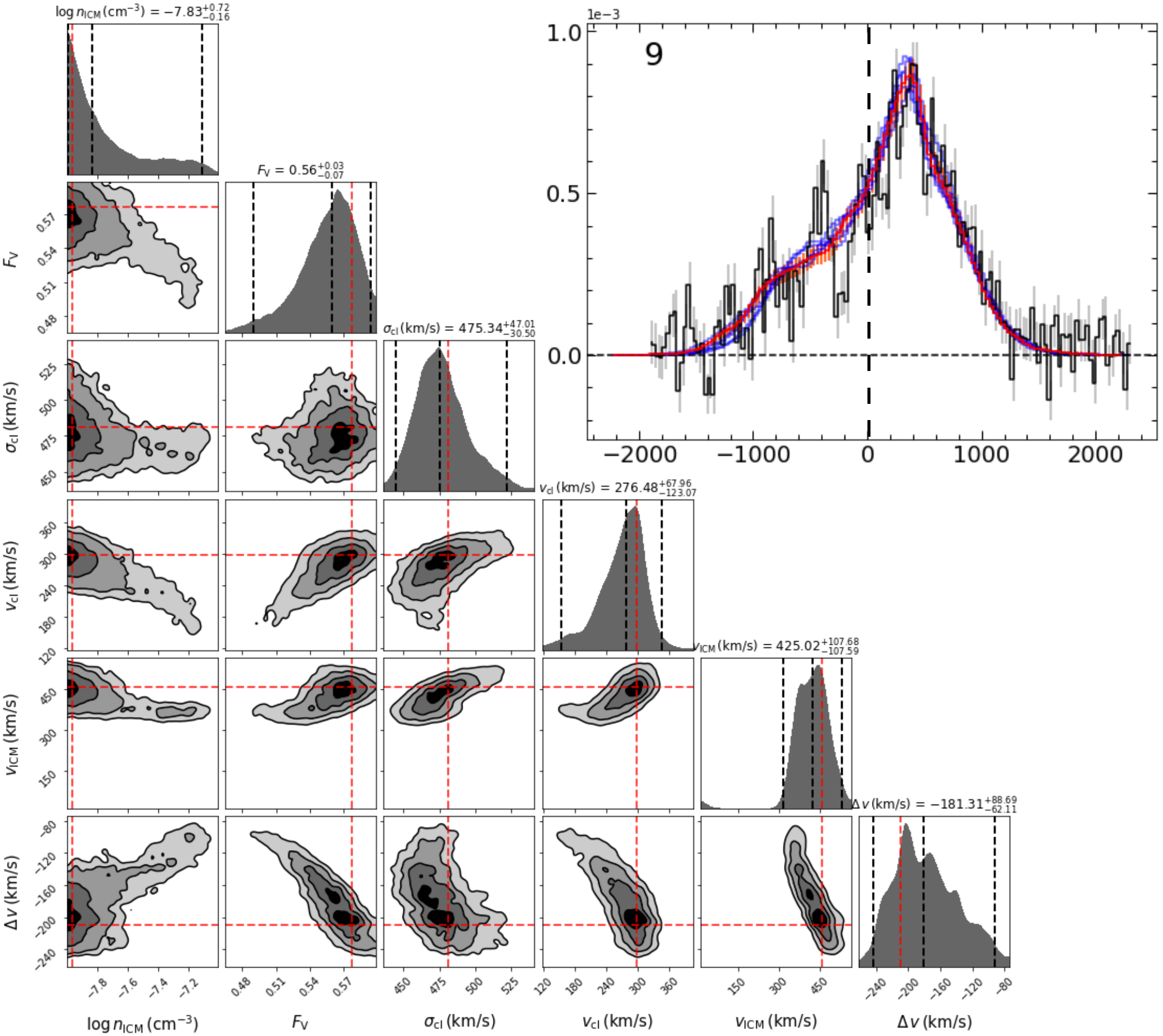}
\end{figure*}

\begin{figure*}
\centering
\includegraphics[width=\textwidth]{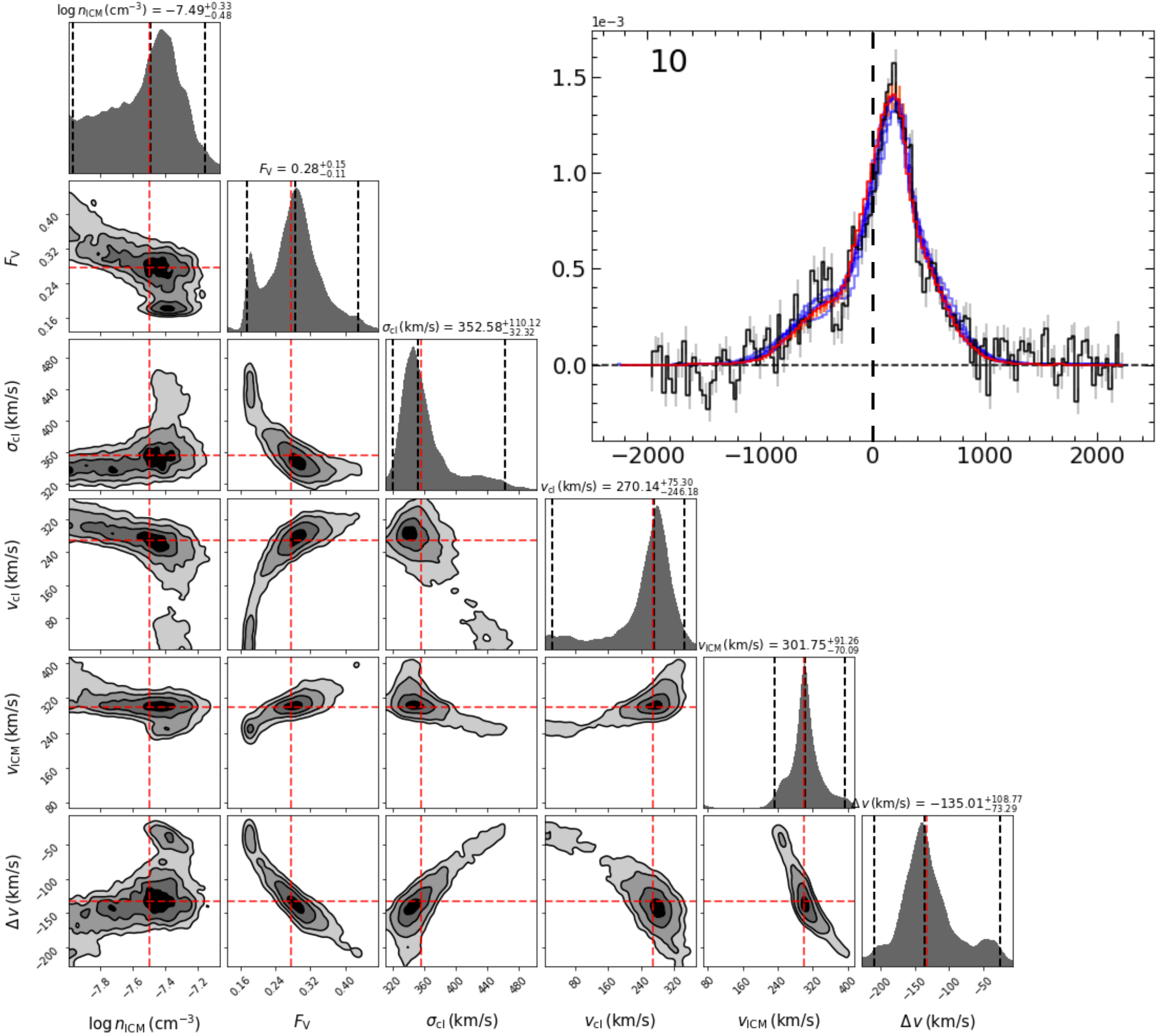}
\end{figure*}

\begin{figure*}
\centering
\includegraphics[width=\textwidth]{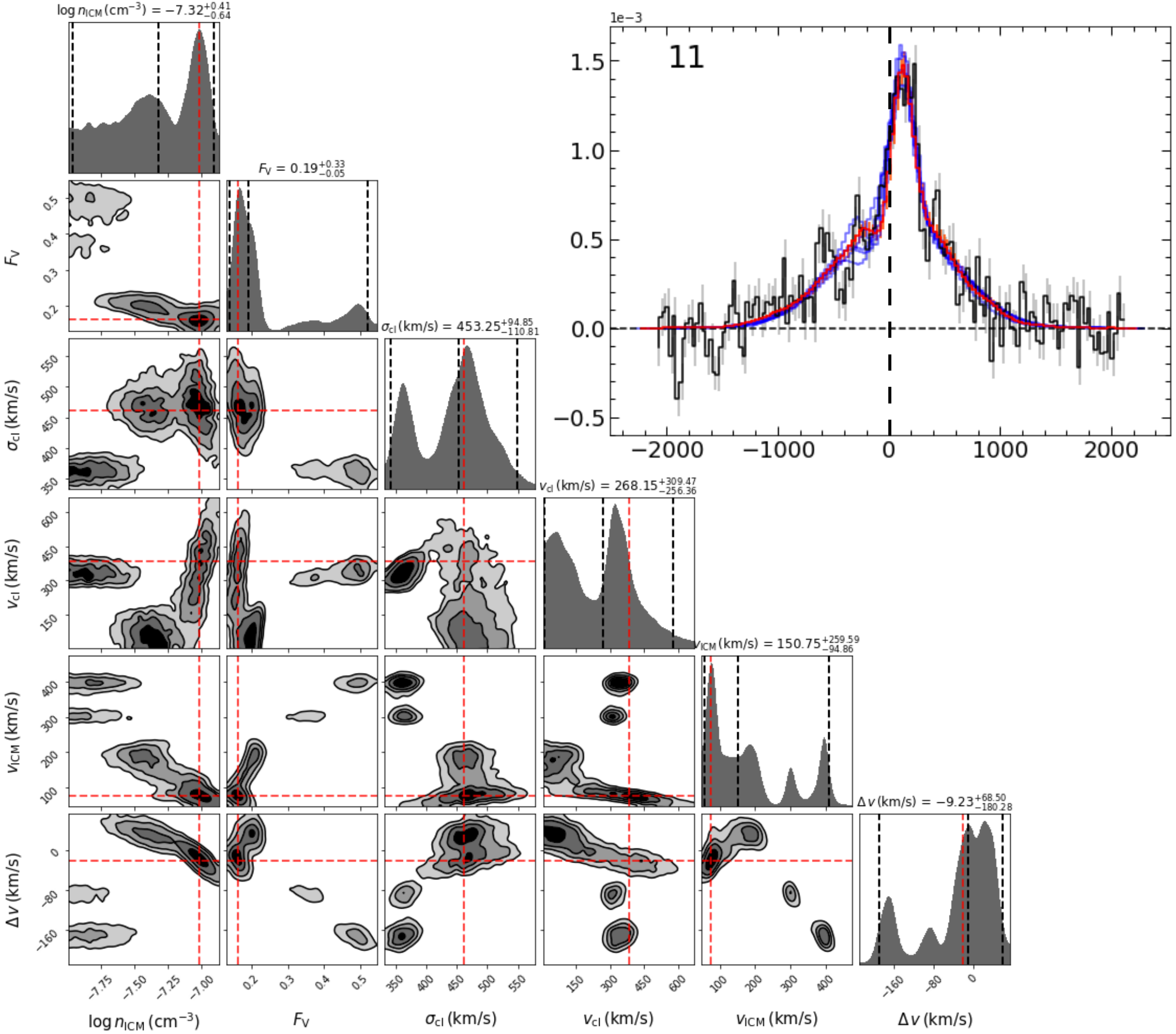}
    \caption{Joint and marginal posterior probability distributions of the multiphase clumpy model parameters for all eleven representative \lya\ spectra derived from nested sampling. The vertical black dashed lines indicate the [2.5\%, 50\%, 97.5\%] quantiles (i.e., 2-$\sigma$ confidence intervals). The vertical red dashed lines indicate the locations of the maximum posterior probability. The upper right panels show the best-fits (red, with orange 1-$\sigma$ Poisson errors), the observed \lya\ profiles (black, with grey 1-$\sigma$ error bars) and five model spectrum samples from nested sampling (blue) in the same way as Figure \ref{fig:Lya_linemaps}.
    \label{fig:posterior}}
\end{figure*}

\end{appendix}

\bsp	
\label{lastpage}
\end{document}